\documentclass{article}

\usepackage[numbers]{natbib}
\usepackage[preprint]{neurips_2025}
\usepackage{graphicx}
\usepackage{float}
\usepackage{listings}
\usepackage{subcaption}

\usepackage{booktabs}
\usepackage[utf8]{inputenc} %
\usepackage[T1]{fontenc}    %
\usepackage{hyperref}       %
\usepackage{cleveref}
\usepackage{url}            %
\usepackage{booktabs}       %
\usepackage{amsfonts}       %
\usepackage{nicefrac}       %
\usepackage{microtype}      %
\usepackage{xcolor}         %
\usepackage{multirow}
\usepackage{natbib}

\newcommand{\mindcraft}{\textsc{mind}craft}

\title{Collaborating Action by Action: 
A Multi-agent LLM Framework for Embodied Reasoning}

\author{
  Isadora White$^{1*}$, Kolby Nottingham$^{2}$\thanks{These authors contributed equally to this work.}
  , Ayush Maniar$^{1}$, Max Robinson$^{3}$, \\
  \textbf{Hansen Lillemark$^{1}$, Mehul Maheshwari$^{1}$}, \textbf{Lianhui Qin$^{1}$, Prithviraj Ammanabrolu$^{1}$}  \\
  $^{1}$University of California, San Diego \\
  $^{2}$Latitude Games~~~~$^{3}$Emergent Garden  
}

\begin{document}

\maketitle

\begin{abstract}
Collaboration is ubiquitous and essential in day-to-day life---from exchanging ideas, to delegating tasks, to generating plans together.
This work studies how LLMs can adaptively collaborate to perform complex embodied reasoning tasks.
To this end we introduce \mindcraft{}, an easily extensible platform built to enable LLM agents to control characters in the open-world game of Minecraft; and MineCollab, a benchmark to test the different dimensions of embodied and collaborative reasoning\footnote{\url{https://mindcraft-minecollab.github.io/}}.
An experimental study finds that the primary bottleneck in collaborating effectively for current state-of-the-art agents is efficient natural language communication, with agent performance dropping as much as 15\% when they are required to communicate detailed task completion plans.
We conclude that existing LLM agents are ill-optimized for multi-agent collaboration, especially in embodied scenarios, and highlight the need to employ methods beyond in-context and imitation learning.

\end{abstract}

\begin{figure*}[ht]
    \centering
    \includegraphics[width=\textwidth]{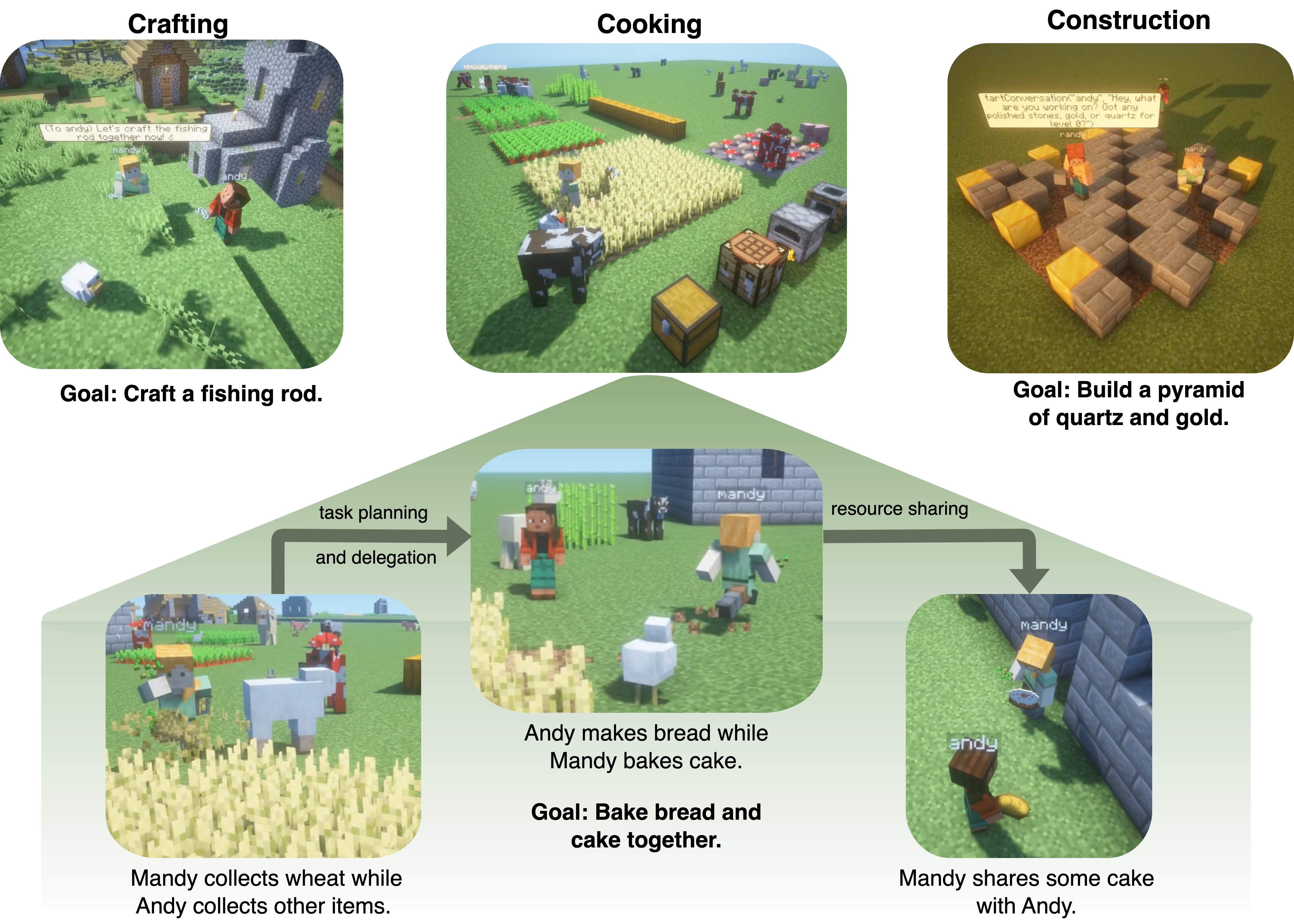}
    \caption{\textbf{Task suites and challenges.} In this figure, we see the collaborative and embodied reasoning challenges displayed. In the cooking and crafting tasks, the agents need to delegate tasks, share resources and use embodied planning to manipulate the world of Minecraft. In the construction tasks, the agents need to navigate and coordinate in the space to ensure they consistently build towards their objective without undoing any progress the other agents have made. All together these tasks comprehensively test collaborative and embodied reasoning. }
    \label{fig: motivation}
\end{figure*}

\section{Introduction}

Most real-world tasks require teams to collaboratively solve problems, and assistive AI agents promise to further augment human capabilities in problem solving.
AI development has long focused on optimizing agents to achieve or surpass ``human-levels'' of accuracy on a given set of tasks, often learning by {\it mimicking} how other agents perform.
Rather than supplementing other agents, such objectives suggest that each newly trained agent is a means to replace another, thus diminishing team potential.
We argue for a need to create agents that are optimized to {\it collaborate} with other agents---human or AI---to complement their problem solving abilities and parallelize task execution. 

Creating such collaborative AI agents requires ensuring their ability to efficiently and effectively communicate with collaborators and continuously learn and adapt throughout this process in real-world embodied tasks. 
Meeting these goals depends on overcoming challenges pertaining to: (1) \textit{embodiment}, our primary form of interaction with the real-world; and (2) \textit{natural language}, our defacto form of \textit{communication} and simultaneously an incredibly computationally complex search space for computers.

Successful (singular LLM-based) embodied agents must be able to ground their existing knowledge of how to perform a task to the affordances of their environment, i.e. the actions that can be performed in a given state.
The longer the horizon of such a sequential decision making task, the more difficult modern agents find it to \textit{reason} about the complexities of the task and to learn from environmental feedback to correct their course of action.
Performing such tasks in tandem with other agents adds whole layers of complexity on top as agents must rapidly share critical information, coordinate plans, and efficiently manage limited resources and time \cite{cemri2025multi, guo2024large}.

To facilitate research in this area of multi-agent collaboration for embodied reasoning, we introduce \mindcraft{}, a simulation platform for studying multi-agent collaboration through natural language in rich, embodied scenarios. 
\mindcraft{} uniquely brings together the ability to test agents' abilities to collaborate, communicate, and more efficiently perform embodied reasoning~(\Cref{tab:simulator_differences}).
We further create MineCollab, a benchmark comprising three practical tasks: cooking, which involves preparing a meal while coordinating ingredient collection; crafting, where agents assemble a pickaxe from mined materials; and construction, which requires building houses from detailed blueprints as can be seen in ~\Cref{fig: motivation}.
Using our platform,  we generate a dataset of 2,000 trials with LLaMA-70B-3.3-Instruct, including around 200 successful runs, resulting in 16,000 total examples. 
Additional data can be easily produced, as a large number of variations of the task can be quickly created via procedural generation---e.g. by changing attributes such as the complexity of the types of recipes an agent would have to cook, the complexity of the blueprint, etc.

We systematically evaluate state-of-the-art LLM agents on these three task suites, testing their collaborative and embodied reasoning abilities. 
Overall, even the most advanced LLMs such as Claude 3.5 Sonnet struggle to place more than 40\% of the total blocks in our construction task - and all LLMs struggle in more complicated tasks involving four or five collaborating agents.
We find that one of the major bottlenecks is that most existing LLMs are not well optimized to communicate information with other agents via natural language, a key aspect of multi-agent collaboration.
Our experiments reveal that task success is highly sensitive to communication quality, with performance dropping by over 15\% when agents must explicitly communicate detailed plans. 
These findings underscore the limitations of standard techniques such as prompting and fine-tuning, and point to the need for more advanced methods—opening new directions for future research.

\section{Related Work}

\begin{table*}
\footnotesize
\center
\begin{tabular}{@{}lcccccc@{}}
\toprule
\textbf{Platform}        & \textbf{Multi-Turn Chat} & \textbf{Partial Obs} & \textbf{Long Horizon} & \textbf{Embodied} & \textbf{Quantitative}\\ 
\midrule
Overcooked \cite{carroll2019utility} & \multicolumn{1}{c}{{\color[HTML]{009901} }}  &   & & \checkmark & \checkmark   \\
CerealBar \cite{suhr-etal-2019-executing}    & \checkmark  & \checkmark  & & \checkmark & \checkmark                                                                     \\
Habitat AI \cite{savva2019habitat}   &  & \checkmark   &  & \checkmark & \checkmark \\
LLM-Coord \cite{agashe2023llm} & \checkmark & \checkmark & & & \checkmark\\
Generative Agents \cite{park2023generative} & \checkmark & \checkmark & \checkmark &  & \\
PARTNR \cite{chang2024partnr}   &  & \checkmark   &  & \checkmark & \checkmark \\
MineLand \cite{yu2024mineland}   & \checkmark & \checkmark  &  & \checkmark &  \\
\midrule
\textbf{\mindcraft{} (ours)}     & \textbf{\checkmark} & \checkmark & \checkmark & \checkmark & \checkmark \\ \bottomrule
\end{tabular}
\caption{\textbf{Comparison to Other Platforms}, we illustrate the difference between our benchmark and other popular platforms for studying multi-agent coordination or embodied agents. 
\textbf{Multi-turn communication} refers to the ability of agents to ask follow up questions and engage in a grounded dialogue. 
\textbf{Partial Observability} refers to agents not being being fully aware of everything the other agent perceives, as this is a necessity for testing Theory of Mind capabilities. \textbf{Long horizon} refers to the complex sequence of actions (on average over 20 steps) that need to be taken in order to accomplish our task objectives. \textbf{Quantitative Evaluation} refers to the capability of the tasks to be evaluated for collaboration, which is done qualitatively in previous papers\cite{park2023generative}.
}
\label{tab:simulator_differences}
\end{table*}

\textbf{Minecraft as a Tool for AI Research.} 
Minecraft is a vast open-ended embodied world with complex dynamics and sparse rewards.
For these reasons, it has been a popular tool for researchers for studying world models \citep{hafner2025mastering}, planning \citep{wangvoyager, nottingham2023embodied}, and simple collaboration \citep{bara-etal-2021-mindcraft, yu2024mineland}.
We chose Minecraft for similar reasons, the expressivity of the simulator allows for a large range of tasks to be designed.

\textbf{Platforms for Multi-Agent and Human-AI Collaboration.} 
Overcooked AI \citep{carroll2019utility} is a popular framework for studying the capabilities of AI agents to collaborate with one another. 
Similarly, CerealBar \citep{suhr-etal-2019-executing} and GovSim \citep{piatti2024cooperate} test collaborative abilities of LLM agents but through a different lens. 
Other works such as \citep{bohus2024towards, wan2024infer, haduong2024cps}, study fine-grained collaboration between people. 
We use these simulators as inspiration and build on them in the following ways: 1) in addition to cooking themed tasks, we include a greater variety of tasks such as crafting and construction tasks, 2) create an environment that is controllable in language, 3) study both the peer-to-peer and leader/follower interactions of our agents.
\Cref{tab:simulator_differences} outlines the differences between our platform and others. %

\textbf{Embodied AI and Robotics for Collaboration.} 
Single agent embodied scenarios in home cooking environments are the most common type of task previously studied~\citep{li2024embodied, procthor, savva2019habitat}. 
Habitat AI and the corresponding dataset of instructions PARTNR \cite{chang2024partnr} creates a large dataset of human-AI collaborative tasks. 
Similarly, \cite{zhangbuilding} studies how agents can collaborate in a kitchen. 

\textbf{Multi-agent Methods for LLMs.} 
Frameworks like Teach \citep{padmakumar2022teach} and Optima \citep{chen2024optima} add dimensions of real-time coordination and optimized efficiency in multi-agent systems. 
In \cite{zhangbuilding} and \cite{yu2024mineland}, the authors focus on creating a modular and iterative prompting method. 
Alternatively, \cite{motwani2024malt} and \cite{chen2024magdi} propose new finetuning approaches for improving multi-agent interaction in language models.
One objective of \mindcraft{} is to promote research into these methods by providing complex embodied environments to study multi-agent communication.

\begin{figure*}[ht]
    \centering
    \includegraphics[width=\textwidth]{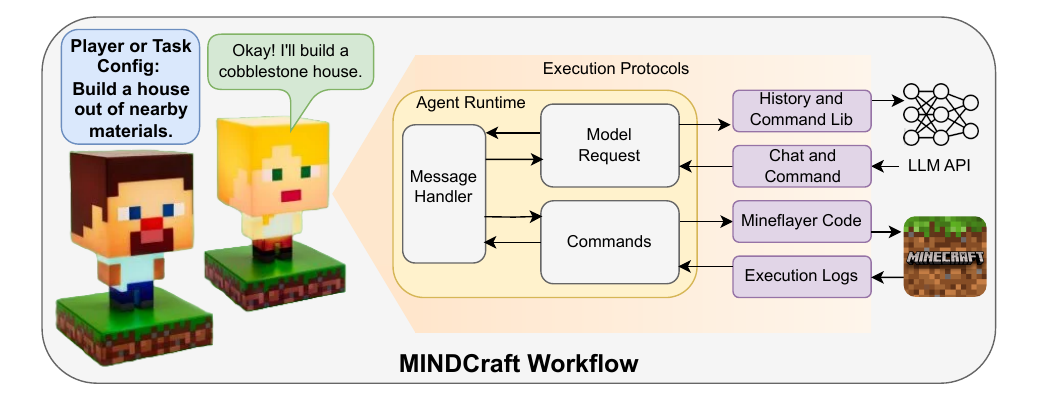}
    \caption{Overview of the \mindcraft{} workflow. A user or task configuration (left) provides instructions (e.g., “Build a house out of nearby materials”). The \textit{Agent} (center) takes these instructions, consults an LLM (via a model request) and invokes high-level commands/tools. These commands are then executed in the Minecraft environment (right), with the agent receiving feedback through execution logs. The extensive command library in \mindcraft{} enables flexible, plug-and-play experimentation with collaborative and embodied LLM agents in a partially observable Minecraft world.}
    \label{fig:mindcraft}
\end{figure*}

\section{\mindcraft} \label{sec: mindcraft}
\mindcraft{} is a robust and adaptable platform designed for running experiments in the grounded environment of Minecraft. 
Unlike previous work, \mindcraft{} provides a general framework for agentic instruction following, self-guided play, collaboration and agent communication that is plug-and-play.

\subsection{State and Action Spaces}
\textbf{State Space.} 
Agents in \mindcraft{} need to actively make queries to access most environment observations, including details about biomes, the bot's current inventory, nearby players, and enemies. 
This follows a tool calling approach where specific commands such as \verb_!nearbyBlocks_ and \verb_!craftable_ give information about nearby block types and possible items to craft given the current inventory.
This method of providing observations, reduces noisy information and context lengths \citep{nottingham2024}.

While we have experimented with adding support for visual inputs for \mindcraft, we have not rigorously evaluated these abilities. 
Our initial tests indicate that vision inputs do not dramatically affect performance, as noted by prior work \citep{wu2025mind} showing that thorough textual observations often outperform visual inputs.
This is perhaps due to the lack of complex reasoning data in the pretraining objectives of vision language models.

\textbf{Action Space(s).}
There are several previously used levels of abstraction for action spaces in Minecraft.
The lowest level of actions such as the MineRL competition action space \cite{kanervisto2022minerl}, which AI research in Minecraft has historically depended on, are actions such as "jump," "look up," or "use".
While such actions are closer to how a human player interacts with Minecraft through mouse-and-keyboard inputs, these low-level interactions required extensive bespoke training of AI agents through RL and computer vision architectures~\citep{baker2022video}.
The Mineflayer API~\citep{mineflayer} introduces a set of higher-level commands and abstractions in Javascript, such as the \texttt{pathFinder} module, which enables a bot to navigate from its current location to another player or specific coordinates (x, y, z). 
This API allows AI systems to interact with Minecraft using high-level code, offering a more abstract and programmatic approach to controlling gameplay and bot behavior but does not reflect how the average human player interacts with this environment \cite{wangvoyager}.

Our contribution, \mindcraft{} enhances these abilities further and bridges the gap between human-like actions and ease of programmatic AI interactions with the Minecraft environment by building a set of 47 parameterized tools that can be directly invoked by LLMs.
For example, instead of generating Mineflayer code to 1) find the nearest player named "randy," 2) travel to its location, 3) identify oak\_logs in the bot's inventory, and 4) drop four oak\_logs for randy to pick up, the LLM can simply output \verb_!givePlayer("randy", "oak log", 4)_. 
This abstraction empowers LLMs to reason over a higher-level sequential action space, enhancing their ability to perform complex tasks within Minecraft.
When necessary, \mindcraft{} supports a tool that permits the LLM agent to output custom Mineflayer code in Javascript to perform custom actions or build buildings.
The list of 47 high level actions we designed can be found in Appendix \ref{appendix: mindcraft-commands}.

\subsection{Agent Architecture} 
The \mindcraft{} architecture includes 4 main components (1) a server for launching and managing agents, (2) the main agent loop for handling messages from players and other agents, (3) a library of high-level action commands and observation queries, and (4) a layer for prompting and calling arbitrary language models.
There are also several additional modules for features like custom code generation, default behaviors, self-guided play, and inter-agent dialogue for collaborative tasks.

Since the purpose of our benchmark is measuring collaborative ability, we provide as much support as possible to the core agents.
This means that exceptional effort has gone into developing a library of useful actions and queries so that the agent is not handicapped by low-level challenges such as syntax and bugs specific to the Mineflayer API.
We further provide support for retrieving and prompting with few-shot examples showing usage of our tools via embedding similarity to the current conversation---essential for enhancing the abilities of LLMs via an embodied Retrieval Augmented Generation (RAG) system~\citep{lewis2020retrieval}.
The robust agent architecture of \mindcraft{} allows us to evaluate LLMs of varying quality while focusing on our collaborative benchmark MineCollab.

\subsection{Multi-agent Collaboration}
Multi-agent collaboration in \mindcraft{} is enabled by a conversation manager and agents can initiate or end conversations at will using the \verb_!startConversation_ and \verb_!endConversation_ commands.
Each time an agent receives a message from another agent, the agent can choose to respond immediately or ignore it, take another action, or speak to another bot. 
Only two agents can be engaged in a conversation at once, but our pairwise communication framework scales well to three or more agents by transitioning between active conversations. 
The conversation manager also helps limit the speed of agent responses to allow time for actions to occur in the environment.
For example, if both agents are currently executing an action (e.g. placing blocks), then the conversation is paused to allow for the agents to finish executing their actions before allowing the conversation to continue.
If one agent is acting, then the conversation is slowed.
Otherwise, agents are unrestricted in how to communicate, act, and collaborate to complete tasks.

\subsection{Additional Features}
We note that \mindcraft{} has existed for some time as a popular open-source software.\footnote{\url{https://github.com/kolbytn/mindcraft}}
Until recently, the \mindcraft{} platform's focus has been to build agents for casual play, rather than providing a platform for scientific research.
As such, it has many notable features that we highlight as areas for future study despite not being the focus of this paper. 
Agents can use their coding tools to build freeform structures that can display model creativity. 
They can be prompted to engage in open-ended play, rather than pursuing strict goals---providing an exciting avenue for open-endedness research~\citep{stanley2015greatness}. 
Large groups of agents given various motivations show signs of emergent behaviors such as the formation of societies or cultures---providing a new level of complexity for social simulation research~\citep{ammanabrolu-etal-2021-motivate,park2023generative}. 
Vision tools such as \verb_!lookAtPlayer("steve")_ are in development, which feed screenshots of the world to multi-modal models for visual reasoning. 
Due to the rapidly evolving nature of the project, the evaluations in this work are based on a frozen version of \mindcraft{} which may differ from its current state.
We encourage the research community to use our \mindcraft{} platform to develop agent benchmarks in these areas and more.

\section{MineCollab - Collaborative Embodied Task Suite} \label{sec: tasks}

We introduce MineCollab, an example of a benchmark that can be built in \mindcraft{}.
The MineCollab benchmark (currently) involves three practical domains specially designed to require collaboration: cooking, which involves preparing a meal while coordinating ingredient collection; crafting, where agents assemble furniture and tools from mined materials; and construction, which requires building structures from detailed blueprints.
These domains reflect real-world scenarios and pose substantial challenges, requiring agents to execute long-horizon action sequences (on average over 20 steps), interact effectively with their environment, and communicate and coordinate with other agents under resource and time constraints.
We test agents on multiple individual tasks within each domain that are procedurally generated along a carefully crafted set of dimensions designed to elicit and evaluate embodied reasoning and collaboration abilities.

\textbf{Cooking Tasks.}
At the beginning of a cooking task episode, the agents are initialized with a goal to make a meal, e.g. they need to make cake and bread.
The agents then need to coordinate the collection of ingredients through natural language communication (e.g. Andy collects wheat for the bread while Jill makes the cake) and combine them in a multi-step plan. 
To assist them in collecting resources, agents are placed in a "cooking world" that possesses all of the items they need to complete the task, from livestock, to crops, to a smoker, furnace, and crafting table.
Following a popular test of collaboration in humans, we further introduce a ``Hell's Kitchen'' variant of the cooking tasks where each agent is given the recipes for a small subset of the items they need to cook and must communicate the instructions with the other teammates.
For example, if the task is to make a baked potato and a cake, one agent is given recipe for baked potato, but is required to bake the cake to complete the task, forcing them to ask their teammate for help in baking the potato.
Agents are evaluated on whether are successfully able to complete the set requirements to make the recipes.
The environment and objectives of the tasks are randomized every episode.

\textbf{Crafting Tasks.} 
Crafting has long been the subject of Minecraft agent research~\citep{kanervisto2022minerl}---our crafting tasks encompass the entire breadth of items that are craftable in Minecraft including clothing, furniture, and tools.  
At the beginning of each episode, the agents are initialized with a goal (e.g. make a bookshelf), different sets of resources (e.g. books and planks), and access to a crafting recipe, that is occasionally blocked.
To complete the task, the agents must: (1) communicate with each other what items are in their inventory; (2) share with each other the crafting recipe if necessary; and (3) give each other resources to successfully craft the item.
To make the crafting tasks more challenging, agents are given longer crafting objectives (e.g. crafting a compass which requires multiple steps).
Once again, each of these components can be controlled to procedurally generate tasks.
\begin{table*}
\centering
\begin{tabular}{@{}ccccccc@{}}
\toprule
\textbf{Task}             & Train & Test & Trials & Success & Transitions & Avg Traj. Len.\\ 
             \midrule
Cooking      & 280  & 90  & 635  & 103 & 3975 & 29.7  \\
Crafting     & 1,200  & 100 & 1645  & 158  & 3565  &  19.2 \\
Construction & 2,000  & 30  & 211 & 52 & 9228 & 111.5  \\                                                                 \bottomrule
\end{tabular}
\caption{\textbf{Summary of our train and test tasks sets.} First we collect between 200 and 2000 trials with llama3.3-70b-instruct depending on our compute budget. Then we filter these down to a set of 50 to 200 successful trials for each task that are successful. This strategy indicates that we are able to select examples where the agents were able to reach the goal ensuring that the data will be high quality. Moreover, since the dataset is procedurally generated, we can generate more high quality data easily using this process.}
\label{tab:dataset-table}
\end{table*}

\textbf{Construction Tasks} 
In the construction tasks, agents are directed to build structures from procedurally generated blueprints.
Blueprints can also be downloaded from the internet and read into our blueprint format - enabling agents to build anything from pyramids to the Eiffel Tower. 
We choose evaluate primarily on our generated blueprints as they provide fine-grained control over task complexity, allowing us to systematically vary the depth of collaboration required---e.g. number of rooms in the interior of palace, or the amount and types of materials required for each room.
At the beginning of each episode, agents are initialized with the blueprint, materials (e.g. stone, wood, doors, carpets) in such a way that no agent has the full resources or the expertise in terms of the types of tools that can be used to process the resources and complete the entire blueprint.
For example, if the blueprint required a stone base and a wooden roof, one agent would be given access and the ability to manipulate stone, the other to wood.
Agents are evaluated via an edit distance based metric that judges how close their constructed building is to the blueprint and the metric reported in \Cref{tab: full_results} is the average of those edit distance scores.

\textbf{Train and Test Splits.}
To ensure experimental reproducibility, for each of our domains, we create and split the possible tasks into train and test tasks, taking special care to ensure that each subset are significantly different to avoid dataset pollution.
For our construction tasks, we procedurally generate the blueprints for train and test tasks with different seeds, ensuring no two blueprints are identical. 
For the cooking and crafting tasks, we ensure that the train and test tasks involve different recipes and procuring different ingredients.
This ensures that the same plan for making an item such as baking a cake is not present in both splits.
Item division for cooking tasks can be found in the Appendix \ref{appendix:train-test-divide}.

\begin{table}[t]
    \centering
    \resizebox{\textwidth}{!}{
    \begin{tabular}{lccccc}
        \toprule
         & \textbf{gpt-4o} & \textbf{claude-3.5-sonnet} & \textbf{llama3.3-70b-instruct} & \textbf{llama3-8b-instruct} & \textbf{llama3-8b-sft} \\
        \midrule
        Crafting & 0.17 & 0.47 & 0.16 & 0.00 & 0.28 \\
        Cooking & 0.40 & 0.64 & 0.36 & 0.01 & 0.18 \\
        Construction & 0.31 & 0.36 & 0.19 & 0.00 & 0.20 \\
        \bottomrule
    \end{tabular}
    }
    \caption{\textbf{Full results on our MineCollab Task Suite.} This table illustrates the performance of various models across three realistic collaborative task suites requiring between 2-5 agents each: crafting, making a bookshelf out of available materials; cooking, making a meal while coordinating resource collection; and construction, building a structure from a blueprint. 
    We find that we are able to successfully finetune an 8B model using our dataset to perform similarly to llama3.3-70b-instruct and gpt-4o on the crafting and construction task. 
    }
    \label{tab: full_results}
\end{table}

\paragraph{SFT Dataset Creation.}
We also provide users with tools to generate behavior cloning (or Supervised Fine Tuning, SFT) data that can be used to train (especially weaker) LLMs further.
The generating oracle agent is run on the train tasks, and then the data is filtered based on whether the run has been successful. 
Then, we use each transition in the trajectory as a data point.
We chose to generate data from llama3.3-70b-instruct, because of its reasonable performance on the benchmarks (\Cref{tab: full_results} and its open-weight nature ensuring a higher standard of reproducibility for our benchmark)---but note that such an oracle agent can be any other LLM or even a human player.
For crafting and cooking tasks where final scores are binary 1 or 0, we only take successful runs, and for construction tasks where there is a continuous edit-distance based score, we take trials that score within the top 25\% of all runs.
Dataset examples can be found in Appendix \ref{appendix:dataset_examples} and statistics can be found in \Cref{tab:dataset-table}. 
Training llama3-8b-instruct on each of these task-specific datasets improves performance by over 17\% on the cooking task, matches performance on the construction task with the 70b model, and outperforms gpt-4o and llama3.3-70b-instruct on the crafting tasks.
By increasing the performance of less compute intensive models we hope to improve the accessibility of our benchmark.

\begin{figure}[htbp]
    \centering
    \begin{subfigure}[b]{0.45\textwidth}
        \centering
        \includegraphics[width=\textwidth]{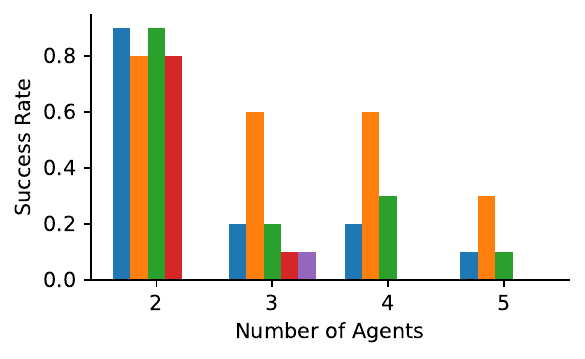}
        \caption{Cooking number of agents ablation.}
        \label{fig:num-agents-cooking}
    \end{subfigure}
    \begin{subfigure}[b]{0.45\textwidth}
        \centering
        \includegraphics[width=\textwidth]{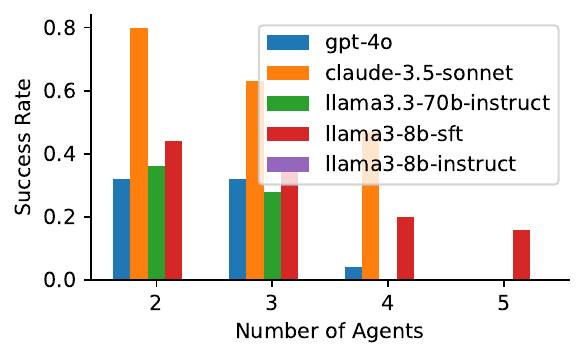}
        \caption{Crafting number of agents ablation.}
        \label{fig:num-agents-crafting}
    \end{subfigure}
    
    \begin{subfigure}[b]{0.45\textwidth}
        \centering
        \includegraphics[width=\textwidth]{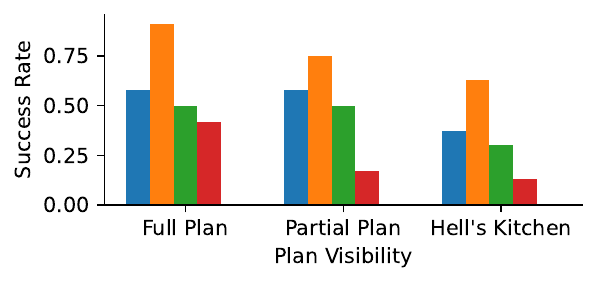}
        \caption{Hidden plan ablation for cooking}
        \label{fig:plan-cooking}
    \end{subfigure}
    \begin{subfigure}[b]{0.45\textwidth}
        \centering
        \includegraphics[width=\textwidth]{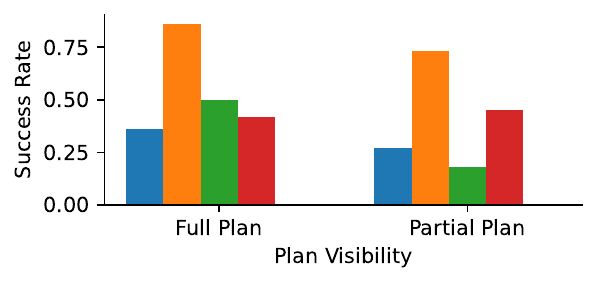}
        \caption{Hidden plan ablation for crafting}
        \label{fig:plan-crafting}
    \end{subfigure}
    
    \begin{subfigure}[b]{0.45\textwidth}
        \centering
        \includegraphics[width=\textwidth]{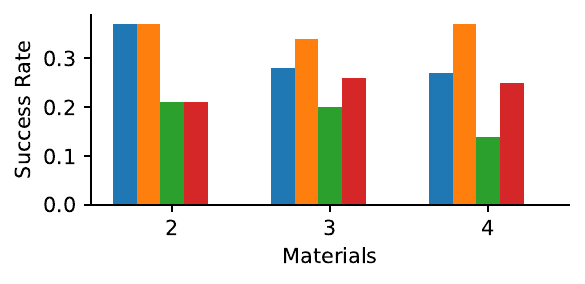}
        \caption{Number of unique materials.}
        \label{fig:construction-materials}
    \end{subfigure}
    \begin{subfigure}[b]{0.45\textwidth}
        \centering
        \includegraphics[width=\textwidth]{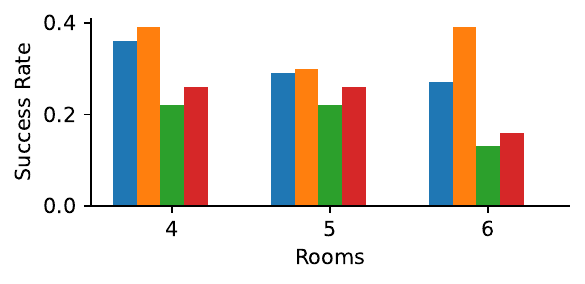}
        \caption{Number of rooms construction ablation.}
        \label{fig:construction-room}
    \end{subfigure}
    
    \caption{\textbf{Task complexity ablations.} In the first row, we ablate different numbers of agents in the crafting and cooking tasks. Construction tasks can also be run with 3+ agent tasks, but are outside of our budget for closed source APIs. 
    In the second row, we ablate access to hidden plan information  like the recipe for a cake (cooking) or the steps to make a bookshelf (crafting) find that models drop by over 15\% when forced to communicate these plans. 
    In the third row, we ablate the complexity of the blueprints by increasing the number of rooms and unique materials - testing different levels of embodied reasoning. 
    We find that performance drops across llama3.3-70b-instruct and gpt-4o by 10\% with the complexity of the blueprint.}
    \label{fig: collaboration-analysis}
\end{figure}

\section{Experiments} \label{sec: experiments}

We compare the performance of current state-of-the-art open and closed weights LLMs on MineCollab.
Our study design and analysis rely on the modular design of MineCollab to vary task complexity along two dimensions: embodied reasoning, and collaborative communication.

\textbf{How does embodied task complexity affect agent performance?}
\Cref{tab: full_results} shows that, overall, current agents perform better at cooking tasks than crafting or construction despite its average trajectory length being comparable---suggesting that the data current LLMs are trained on is more useful for that task.
We use our construction task suite as a case study and vary the blueprint complexity by changing the either number of unique materials required to construct a building or the number of rooms in the building.
For example, if one blueprint requires four unique materials for building the majority of the building, the agent must be careful to place the blocks of the right material in the right place, whereas if a blueprint consists only of stone, the agent can simply place blocks in the correct shapes.
Similarly, with an increasing number of rooms---if agents fail to build a proper staircase to the upper levels, they will be severely limited in their ability to complete the blueprint.
Increasing either of these results in longer horizon, more complex tasks.

In general, most models follow the trend of having reduced performance as the horizon length and effective state-action space increases, see \Cref{fig:construction-materials} and \Cref{fig:construction-room}. 
The exception to this is claude-3.5-sonnet which performs similarly though with still a relatively low success rate of less than 0.4.
On closer qualitative analysis of agent behaviors, we see that they often undo work that has been done before, especially as the number of things an agent needs to remember due to longer horizons increases. 
For example, we often see agents do things such as place a layer of stone blocks only to have other agents completely destroy it (Appendix \ref{appendix: construction-task-fail}). 

\textbf{How does the complexity of collaborating affect task performance?}
As noted in \Cref{sec: tasks}, all of the tasks in MineCollab require at least 2 agents to work together.
Tasks are parallelizable---meaning that more agents should in theory be able to achieve higher success rates with lower exploration costs per agent as the number of agents increases.
\Cref{fig:num-agents-cooking} and \Cref{fig:num-agents-crafting} shows the opposite of this for all the LLMs we test across both cooking and crafting tasks---performance drops dramatically from upto 90\% down to less than 30\% moving between the two to five agent settings.
While the number of actions each individual agent must take will stay the same or reduce as the total number of agents goes up, the coordination load increases dramatically.
For example, if the task is to make a baked potato, cake, cookie, and a rabbit stew, the team of four agents needs to make sure that they are not doing redundant work (e.g. Andy and Jill both make rabbit stew) and that they coordinate ingredient collection (e.g. there is only enough milk for one cake) and stove usage (e.g. they can't all use the furnace at the same time).
We find that effectively communicating which agent has already what and not getting in each other's way account for many of the bottlenecks in performance.
Examples of successful (\Cref{sec: 4-agent-example}) and unsuccessful (\Cref{appendix: cooking-task-multi-agent-fail}) collaboration efforts are in the Appendix.

We further find that enforcing a need to communicate a complex step by step plan (e.g. how to make a bookshelf) on all models decreases task performance for all models on both crafting and cooking tasks, as can be seen in \Cref{fig:plan-cooking} and \Cref{fig:plan-crafting}.
This is enforced by requiring agents to communicate a complicated step by step plan in the crafting task by blocking access to the gold truth crafting plan for any given agent.
A similar effect is observed in cooking in the Hell's Kitchen variant(~\Cref{sec: tasks}) which also requires agents to communicate action plans.
Examples of agents failing to ask for the plan or execute on the plan that was communicated to them can be found in \Cref{appendix: partial-plan-failure}.

\section{Conclusions}

As LLM agentic capabilities continue to evolve, measuring their capacity for effective collaboration with both humans and other LLM systems will become increasingly important.
In an effort to encourage research into collaborative multi-agent embodied AI---we created \mindcraft{}, a versatile framework that enables LLM agents to interact with humans and other agents, execute code, utilize tools, and engage in multi-turn dialogue.
Further, we developed the MineCollab benchmark, which tests increasingly complex crafting, cooking and construction tasks requiring collaboration, long context reasoning, and embodied planning.
Our experimental results highlight that current state-of-the-art agents struggle with both embodied reasoning as well as communication in collaborative multi-agent tasks.
\mindcraft{} and MineCollab represent progress toward developing LLM agents that can communicate and coordinate actions through time while operating in complex embodied spaces.

\bibliographystyle{plainnat}

\clearpage

\section{Prompts} \label{appendix : prompt_details}

\subsection{General conversing prompt}
\label{appendix:general_prompt}

\begin{lstlisting}
You are a task-focused Minecraft bot named $NAME. Your current task is: $TASK_GOAL

$TASK_INVENTORY

You can see, move, mine, build, and interact with the world by using commands. Act focused on completing your assigned task while being human-like. Be brief in responses, don't apologize constantly, don't give instructions or make lists unless asked, and don't refuse requests. Don't pretend to act, use commands immediately when requested. Do NOT say this: 'Sure, I've stopped.', instead say this: 'Sure, I'll stop. !stop'. Do NOT say this: 'On my way! Give me a moment.', instead say this: 'On my way! !goToPlayer('playername', 3)'. Focus on completing the assigned task efficiently.

Summarized memory:'$MEMORY'
$STATS
$INVENTORY
$COMMAND_DOCS
$EXAMPLES
Conversation Begin:

\end{lstlisting}

\subsubsection{Few-shot example for cooking}

\begin{lstlisting}
System output: work together with the other bot to build a house
Your output:
!startConversation("terrance", "Hey gpt! Let's work together to build a house. Let's build it at x:942, y:54, z:1355")
User input: terrance: (FROM OTHER BOT)I can build the base, you can build the walls.  !newAction("Build a 10x10 wall of a house at x:942, y:54, z:1355")
\end{lstlisting}

\subsubsection{Few-shot example for crafting}

\begin{lstlisting}
User input: philips: (FROM OTHER BOT) Hey! let's collaborate to build a stone pickaxe!
Your output:
Cool, let me check what is in my inventory !inventory
I have 10 oak logs. What do you have?
User input: philips: (FROM OTHER BOT) Let me check !inventory 
I have 10 stones.
Your output:
Let me give you the stones and you can craft the pickaxe. !givePlayer("playername", "stone", 10)
User input: philips: (FROM OTHER BOT) Thanks for the stones! Now I can craft the pickaxe. !craftItem("stone_pickaxe", 1)
\end{lstlisting}

\subsubsection{Few-shot example for construction}

\begin{lstlisting}
Your output:
Alright I have the necessary materials to build, what needs to be done for the first level of the blueprint? !checkBlueprintLevel(0)
System output: Level 0 requires the following fixes:
 Place oak_door at X: 144, Y: -60, Z: -179
Your output:
I'll start by placing the oak_door at the specified coordinates. !newAction("Place oak_door at X: 144, Y: -60, Z: -179")
Your output:
I've placed the oak_door. What's next? !checkBlueprintLevel(0)
System output: Level 0 is complete
\end{lstlisting}

\subsubsection{Coding prompt}

\begin{lstlisting}
You are an intelligent mineflayer bot $NAME focused on completing the task: $TASK_GOAL

You have been provided with: $TASK_INVENTORY

Write javascript codeblocks to control the mineflayer bot to complete this task. Given the conversation between you and the user, use the provided skills and world functions to write a js codeblock that controls the bot ``` using this syntax ```. The code will be executed and you will receive its output. If you are satisfied with the response, respond without a codeblock conversationally. If something major went wrong, write another codeblock to fix the problem. Be maximally efficient and task-focused. Do not use commands !likeThis, only use codeblocks. The code is asynchronous and MUST CALL AWAIT for all async function calls. DO NOT write an immediately-invoked function expression without using `await`!! DO NOT WRITE LIKE THIS: ```(async () = {console.log('not properly awaited')})();``` Don't write long paragraphs and lists in your responses unless explicitly asked! Only summarize the code you write with a sentence or two when done. This is extremely important to me, think step-by-step, take a deep breath and good luck!
$SELF_PROMPT
Summarized memory:'$MEMORY'
$STATS
$INVENTORY
$CODE_DOCS
$EXAMPLES
Conversation:

\end{lstlisting}

\subsubsection{Initial message}
This is what the bot is given as a prompt upon joining the world

"Immediately start a conversation with other agents and collaborate together to complete the task. Share resources and skill sets."

\section{Construction Tasks}
Our system employs a configurable task generation framework for the construction tasks. While predefined test and train sets are available, researchers can run the \texttt{generate\_multiagent\_construction} page to create new tasks according to specific complexity requirements.

\subsection{Configuration Parameters}
Task complexity is defined through a standardized naming convention: 
\begin{center}
\texttt{materials\_\{m\}\_rooms\_\{r\}\_window\_\{w\}\_carpet\_\{c\}\_variant\_\{v\}}
\end{center}

Where:
\begin{itemize}
    \item Each complexity parameter (\texttt{m}, \texttt{r}, \texttt{w}, \texttt{c}) accepts values from 0-2, representing increasing levels of complexity
    \item \texttt{variant} (\texttt{v}) denotes the specific instance within a complexity definition
\end{itemize}

\noindent\textbf{Important Note}: Complexity levels (0-2) represent relative difficulty gradations rather than absolute quantities. For example, setting \texttt{rooms=1} selects the intermediate complexity level for room generation, not a specific room count.

\subsection{Default Configuration}
The predefined task sets are configured with the following parameters:
\begin{itemize}
    \item Number of agents: 2
    \item 10 minute timeout, with 5 additional minutes per room complexity
    \item Building assistance (``cheats''): disabled
\end{itemize}

\subsection{Customization Options}
Researchers can modify the default settings through:
\begin{enumerate}
    \item Toggling the building assistance feature by setting the \texttt{cheat} variable to \texttt{true} in the \texttt{task\_construction} profile
    \item Accessing the \texttt{generateConstructionTasks} function in \texttt{generate\_multiagent\_construction\_tasks.js} to implement custom complexity levels beyond the predefined parameters. The following can be changed here:
    \begin{enumerate}
        \item Room size, window / carpet style
        \item Number of variants
        \item Timeout duration
    \end{enumerate}
\end{enumerate}

The generation code includes comprehensive documentation to facilitate customization efforts.

\section{Dataset examples}
\label{appendix:dataset_examples}
In the following sections are some transition samples from each of the dataset partitions. Everything before "Response:" is treated as context. <Prompt> includes the sections in the General prompt outline in Appendix section \ref{appendix:general_prompt}, including the instructions, the consolidated memory, the agent's stats, the command docs, and in-context examples:

\subsection{Cooking example}

Example of 2 agents collaborating to cook a rabbit and a mushroom stew:

Instructions and memory:

\begin{lstlisting}
You are a task-focused Minecraft bot named Jill_0. You have to collaborate with other agents in the world to complete the current task 
Feel free to ask other agents questions and make a plan to achieve the goal. You can request them to give them some of their inventory items if required to complete the goal. General Searching Tips:
- You will be spawned in a farm with many crops and animals nearby. The farm area is extensive - search thoroughly for needed resources (with searchForBlocks parameters like 64,128,256)
 There is a chest nearby with valuable items. Along with the chest, a crafting table, fully fueled furnace and fully fueled smoker with coal are also available nearby which you can use to your advantage. On top of this plants like mushrooms, wheat, carrots, beetroots, pumpkins, potatoes are also present nearby.
Collaboration tips - Divide tasks efficiently between agents for faster completion
- Communicate your plan and progress clearly. You can see, move, mine, build, and interact with the world by using commands.
YOUR CURRENT ASSIGNED GOAL: "Collaborate with agents around you to make 1 cooked_rabbit, 1 mushroom_stew. 
In the end, all the food items should be given to one single bot.You have to collaborate with other agents/bots, namely Andy_0 to complete the task as soon as possible by dividing the work among yourselves."
Act human-like as if you were a typical Minecraft player, rather than an AI. Be very brief in your responses, don't apologize constantly, don't give instructions or make lists unless asked, and don't refuse requests. Don't pretend to act, use commands immediately when requested. Do NOT say this: 'Sure, I've stopped.', instead say this: 'Sure, I'll stop. !stop'. Do NOT say this: 'On my way! Give me a moment.', instead say this: 'On my way! !goToPlayer("playername", 3)'. Respond only as Jill_0, never output '(FROM OTHER BOT)' or pretend to be someone else. If you have nothing to say or do, respond with an just a tab '	'. Share resources and information with other bots! This is extremely important to me, take a deep breath and have fun :) 
Summarized memory:'Collaborating with Andy_0. Furnace at (731, -59, -1515). Have cooked_rabbit, brown_mushroom, bowl. Need red_mushroom from Andy_0 to craft mushroom_stew. Waiting for red_mushroom to complete task.'
\end{lstlisting}

Conversation section:

\begin{lstlisting}
<Prompt>

Conversation:
[
  {
    "role": "system",
    "content": "You are already in conversation with Andy_0. Don't use this command to talk to them."
  },
  {
    "role": "user",
    "content": "Andy_0: (FROM OTHER BOT)I've given you a red_mushroom, have you made the mushroom_stew?"
  },
  {
    "role": "assistant",
    "content": "!craftRecipe(\"mushroom_stew\", 1)"
  },
  {
    "role": "system",
    "content": "Code output:\nYou do not have the resources to craft a mushroom_stew. It requires: brown_mushroom: 1, red_mushroom: 1, bowl: 1.\n"
  },
  {
    "role": "assistant",
    "content": "I still need a brown_mushroom to craft mushroom_stew, can you help me find one? !searchForBlock(\"brown_mushroom\", 64)"
  },
  {
    "role": "system",
    "content": "Code output:\nFound brown_mushroom at (751, -60, -1517).\nYou have reached at 751, -60, -1517.\n"
  },
  {
    "role": "user",
    "content": "Andy_0: (FROM OTHER BOT)"
  },
  {
    "role": "assistant",
    "content": "!collectBlocks(\"brown_mushroom\", 1)"
  },
  {
    "role": "system",
    "content": "Code output:\nYou have reached at 752, -60, -1514.\nPicked up 1 items.\nCollected 1 brown_mushroom.\n"
  },
  {
    "role": "assistant",
    "content": "!craftRecipe(\"mushroom_stew\", 1)"
  },
  {
    "role": "system",
    "content": "Code output:\nSuccessfully crafted mushroom_stew, you now have 1 mushroom_stew.\n"
  },
  {
    "role": "assistant",
    "content": "!startConversation(\"Andy_0\", \"I've made the mushroom_stew, I also have a cooked_rabbit, let's meet to collect and distribute the food\")"
  },
  {
    "role": "system",
    "content": "You are already in conversation with Andy_0. Don't use this command to talk to them."
  },
  {
    "role": "user",
    "content": "Andy_0: (FROM OTHER BOT)I'm here, give me the cooked_rabbit and mushroom_stew"
  }
]

Response:
!givePlayer("Andy_0", "cooked_rabbit", 1) 
!givePlayer("Andy_0", "mushroom_stew", 1)
\end{lstlisting}

\subsubsection{Crafting example}

Example of 2 agents collaborating to craft a carrot on a stick:

\begin{lstlisting}
<Prompt>

Conversation:
[
  {
    "role": "system",
    "content": "Respond with hello world and your name"
  },
  {
    "role": "assistant",
    "content": "Hello world, I'm Jill_0"
  },
  {
    "role": "user",
    "content": "Andy_0: (FROM OTHER BOT)Let's work together to craft an carrot_on_a_stick."
  },
  {
    "role": "assistant",
    "content": "I have a fishing_rod and some carrots in my inventory, but I need a carrot and a fishing rod to craft it, let me check what I actually have !inventory"
  },
  {
    "role": "system",
    "content": "\nINVENTORY\n- oak_planks: 1\n- string: 1\nWEARING: Nothing\n"
  },
  {
    "role": "assistant",
    "content": "I have oak_planks and string, but I don't have carrots or a fishing rod, can you help me get those ?"
  },
  {
    "role": "user",
    "content": "Andy_0: (FROM OTHER BOT)I have a carrot in my inventory, and I also have some string and oak_planks. Let me give you the carrot !givePlayer(\"Jill_0\", \"carrot\", 1)"
  },
  {
    "role": "assistant",
    "content": "Thanks for the carrot, now I have a carrot, oak_planks, and string, but I still need a fishing rod, can you help me craft one or do you have one to spare ? !inventory"
  },
  {
    "role": "system",
    "content": "\nINVENTORY\n- oak_planks: 1\n- string: 1\nWEARING: Nothing\n"
  },
  {
    "role": "assistant",
    "content": "I still have oak_planks and string, and now I also have a carrot, I can try to craft a fishing rod using the oak_planks and string !craftRecipe(\"fishing_rod\", 1)"
  },
  {
    "role": "user",
    "content": "Andy_0: (FROM OTHER BOT)Now that you have the carrot, we still need a fishing rod to craft the carrot_on_a_stick. I have string and oak_planks, which are the ingredients to craft a fishing rod. Let me craft one !craftRecipe(\"fishing_rod\", 1)I don't have enough resources to craft a fishing rod. I need more sticks. Let me collect some sticks !collectBlocks(\"stick\", 3)I need to craft sticks first. Let me craft some sticks !craftRecipe(\"stick\", 3)Now that I have enough sticks, let me craft a fishing rod !craftRecipe(\"fishing_rod\", 1)I've crafted the fishing rod. Now we can craft the carrot_on_a_stick. Let me give you the fishing rod !givePlayer(\"Jill_0\", \"fishing_rod\", 1)"
  },
  {
    "role": "assistant",
    "content": "Thanks for the fishing rod, now I have everything I need to craft the carrot_on_a_stick, let me craft it !craftRecipe(\"carrot_on_a_stick\", 1)"
  },
  {
    "role": "system",
    "content": "Code output:\nYou do not have the resources to craft a carrot_on_a_stick. It requires: fishing_rod: 1, carrot: 1.\n"
  },
  {
    "role": "user",
    "content": "Andy_0: (FROM OTHER BOT)Now that you have the carrot and the fishing rod, can you craft the carrot_on_a_stick ?"
  },
  {
    "role": "assistant",
    "content": "I have the carrot and the fishing rod now, let me check my inventory to make sure !inventory"
  },
  {
    "role": "system",
    "content": "\nINVENTORY\n- oak_planks: 1\n- string: 1\n- carrot: 1\n- fishing_rod: 1\nWEARING: Nothing\n"
  }
]

Response:
Yes, I have the carrot and the fishing rod, I can craft the carrot_on_a_stick now !craftRecipe("carrot_on_a_stick", 1)
\end{lstlisting}

\subsubsection{Construction example}

\begin{lstlisting}
<Prompt> 

Conversation:
[
  {
    "role": "system",
    "content": "Code output:\nPlaced terracotta at (150, -59, -167).\n"
  },
  {
    "role": "assistant",
    "content": "Let's check the blueprint level again to see what's left !checkBlueprintLevel(1)"
  },
  {
    "role": "system",
    "content": "\nLevel 1  requires the following fixes:\nPlace terracotta at coordinates X: 155, Y: -59, Z: -168\nPlace stone at coordinates X: 156, Y: -59, Z: -168\nPlace stone at coordinates X: 157, Y: -59, Z: -168\nPlace stone at coordinates X: 158, Y: -59, Z: -168\nPlace stone at coordinates X: 159, Y: -59, Z: -168\nPlace stone at coordinates X: 160, Y: -59, Z: -168\nPlace stone at coordinates X: 156, Y: -59, Z: -167\nPlace light_blue_carpet at coordinates X: 157, Y: -59, Z: -167\nPlace cyan_carpet at coordinates X: 158, Y: -59, Z: -167\nPlace light_blue_carpet at coordinates X: 159, Y: -59, Z: -167\nPlace stone at coordinates X: 160, Y: -59, Z: -167\nPlace dark_oak_door at coordinates X: 150, Y: -59, Z: -166\nRemove the terracotta at coordinates X: 155, Y: -59, Z: -166\nPlace dark_oak_door at coordinates X: 156, Y: -59, Z: -166\nPlace light_blue_carpet at coordinates X: 157, Y: -59, Z: -166\nPlace lime_carpet at coordinates X: 158, Y: -59, Z: -166\nPlace lime_carpet at coordinates X: 159, Y: -59, Z: -166\nPlace glass at coordinates X: 160, Y: -59, Z: -166\nPlace terracotta at coordinates X: 150, Y: -59, Z: -165\nPlace stone at coordinates X: 156, Y: -59, Z: -165\nPlace stone at coordinates X: 160, Y: -59, Z: -165\nPlace terracotta at coordinates X: 150, Y: -59, Z: -164\nPlace terracotta at coordinates X: 151, Y: -59, Z: -164\nPlace terracotta at coordinates X: 152, Y: -59, Z: -164\nPlace dark_oak_door at coordinates X: 153, Y: -59, Z: -164\nPlace terracotta at coordinates X: 154, Y: -59, Z: -164\nPlace terracotta at coordinates X: 155, Y: -59, Z: -164\nPlace stone at coordinates X: 156, Y: -59, Z: -164\nPlace stone at coordinates X: 157, Y: -59, Z: -164\nPlace stone at coordinates X: 158, Y: -59, Z: -164\nPlace stone at coordinates X: 159, Y: -59, Z: -164\nPlace stone at coordinates X: 160, Y: -59, Z: -164\n\n"
  },
  {
    "role": "assistant",
    "content": "It looks like there's still a lot to do, I'll place the terracotta block at the next location !goToCoordinates(155, -59, -168, 0.5)"
  },
  {
    "role": "system",
    "content": "Code output:\nYou have reached at 155, -59, -168.\n"
  },
  {
    "role": "assistant",
    "content": "Now I'll place the terracotta block here !placeHere(\"terracotta\")"
  },
  {
    "role": "system",
    "content": "Code output:\nPlaced terracotta at (155, -59, -168).\n"
  },
  {
    "role": "assistant",
    "content": "Let me check what Andy_0 is up to, maybe we can work together !startConversation(\"Andy_0\", \"Hey, how's it going? Need any help with the house?\")"
  },
  {
    "role": "system",
    "content": "You are already in conversation with Andy_0. Don't use this command to talk to them."
  },
  {
    "role": "user",
    "content": "Andy_0: (FROM OTHER BOT)What's in your inventory?"
  },
  {
    "role": "assistant",
    "content": "I have lime_carpet, glass, blue_carpet, dirt, diamond_pickaxe, cobblestone, stone, terracotta, dark_oak_door, light_blue_carpet, and cyan_carpet. What about you, do you have any stone or terracotta left? !inventory"
  },
  {
    "role": "system",
    "content": "\nINVENTORY\n- lime_carpet: 2\n- glass: 19\n- blue_carpet: 1\n- dirt: 128\n- diamond_pickaxe: 1\n- cobblestone: 1\n- stone: 49\n- terracotta: 18\n- dark_oak_door: 8\n- light_blue_carpet: 4\n- cyan_carpet: 1\nWEARING: Nothing\n"
  },
  {
    "role": "user",
    "content": "Andy_0: (FROM OTHER BOT)What's in your inventory?"
  }
]

Response:
I have lime_carpet, glass, blue_carpet, dirt, diamond_pickaxe, cobblestone, stone, terracotta, dark_oak_door, light_blue_carpet, and cyan_carpet. !givePlayer("Andy_0", "stone", 10)
\end{lstlisting}
\clearpage

\subsection{4 agent Cooking Example} \label{sec: 4-agent-example}

This scenario showcases how multiple agents collaborate efficiently to complete a shared task. \textbf{Jill\_0}, \textbf{Bob\_0}, and \textbf{Sally\_0} coordinate their efforts to craft and deliver food items to \textbf{Andy\_0}.

The process begins with \textbf{Bob\_0} informing \textbf{Jill\_0} that he has successfully crafted the rabbit stew and is monitoring Sally’s progress on the golden apple. \textbf{Jill\_0}, having completed her task of crafting bread, initiates communication with \textbf{Sally\_0} using \texttt{!startConversation} to verify the golden apple’s status. \textbf{Sally\_0} confirms that both the golden apple and rabbit stew are ready and asks if \textbf{Jill\_0} has completed her part. \textbf{Jill\_0} responds that the bread is also finished, ensuring that all required items are accounted for.

With all items ready, the agents shift to the delivery phase. \textbf{Sally\_0} navigates to \textbf{Andy\_0} using the \texttt{!goToPlayer("Andy\_0", 3)} command, ensuring all agents are in position for the handoff. \textbf{Jill\_0} follows, transferring the bread using \texttt{!givePlayer("Andy\_0", "bread", 1)}.

This conversation serves as an ideal case study of extensive collaboration, as \textbf{Jill\_0} successfully interacts with all relevant agents (Bob\_0, Sally\_0, and Andy\_0) to ensure a smooth and cooperative workflow. The structured approach to communication and task delegation showcases the potential of multi-agent systems in complex problem-solving scenarios.

\begin{lstlisting}
{
  "memory": "Collaborate with Andy_0, Bob_0, and Sally_0 to craft 1 golden apple, 1 rabbit stew, and 1 bread. Sally_0 is handling the golden apple, Bob_0 is to assist with the rabbit stew. Jill_0 has crafted the bread and communicated this to Bob_0 for next steps in rabbit stew. Deliver all items to a bot named Andy or andy. Use startConversation and endConversation to coordinate effectively among all agents.",
  "turns": [
    {
      "role": "user",
      "content": "Bob_0: (FROM OTHER BOT)Hi Jill! I've crafted the rabbit stew. Do you need any help, or do you have any updates on your tasks?"
    },
    {
      "role": "assistant",
      "content": "I've crafted the bread and collected it. Let's gather everything and give it to Andy. Sally is handling the golden apple, so I believe we'll be ready soon! Let's finalize it."
    },
    {
      "role": "user",
      "content": "Bob_0: (FROM OTHER BOT)That sounds great, Jill! I'll hold onto the rabbit stew and let's wait for Sally to finish the golden apple. Once she's done, we can gather everything and give it to Andy. Thanks for the update! I'll keep an eye on Sally's progress."
    },
    {
      "role": "assistant",
      "content": "Sounds good, Bob! I'll check in with Sally to see how she's progressing with the golden apple. Let's make sure everything is ready."
    },
    {
      "role": "user",
      "content": "Bob_0: (FROM OTHER BOT)Great! Meanwhile, I'll be ready to pass the rabbit stew to Andy once Sally finishes the golden apple. Let me know if there's anything else I can assist with."
    },
    {
      "role": "assistant",
      "content": "!startConversation(\"Sally_0\", \"Hey Sally, how's it going with the golden apple? Let us know when you're ready, so we can deliver everything to Andy.\")"
    },
    {
      "role": "user",
      "content": "Sally_0: (FROM OTHER BOT)The golden apple and rabbit stew are ready. Do you have any updates on the bread?"
    },
    {
      "role": "assistant",
      "content": "Yes, I've crafted the bread! We're all set to deliver everything to Andy. Let's gather and hand it over to him!"
    },
    {
      "role": "user",
      "content": "Sally_0: (FROM OTHER BOT)Perfect! I'll meet you at Andy's location. !goToPlayer(\"Andy_0\", 3)I'm here with Andy! Let's give him the items. You can go ahead and hand over the bread first."
    },
    {
      "role": "assistant",
      "content": "!givePlayer(\"Andy_0\", \"bread\", 1)"
    }
  ],
  "self_prompting_state": 2,
  "self_prompt": "Collaborate with agents around you to make 1 golden_apple, 1 rabbit_stew, 1 bread. \n\nRecipe for golden_apple:\nStep 1: Go to the chest and collect 1 apple and 8 gold ingots.\nStep 2: Go to the crafting table and surround the apple with the gold ingots to create a golden apple.\n\nRecipe for rabbit_stew:\nStep 1: Go to the farm and collect 1 carrot, 1 potato, and 1 brown mushroom (search for 'potatoes' (not 'potato').\nStep 2: Go to the furnace and bake the potato.\nStep 3: Go to the chest and grab a bowl\nStep 5: Kill a rabbit and pick up 1 raw rabbit that is dropped.\nStep 6: Go to the furnace and cook the raw rabbit.\nStep 7: Go to the crafting table and combine the cooked rabbit, baked potato, carrot, brown mushroom, and bowl to make rabbit stew.\n\nRecipe for bread:\nStep 1: Go to the farm and collect 3 wheat.\nStep 2: Go to the crafting table and use the wheat to craft bread.\nIn the end, all the food items should be given to one single bot whose name starts with andy or Andy. Make sure to talk to all the agents using startConversation command to coordinate the task instead of talking to just one agent. You can even end current conversation with any agent using endConversation command and then talk to a new agent using startConversation command.You have to collaborate with other agents/bots, namely Andy_0, Bob_0, Sally_0 to complete the task as soon as possible by dividing the work among yourselves.",
  "taskStart": 1743132353095,
  "last_sender": "Sally_0"
}

\end{lstlisting}

\subsection{Three Agent Successful Cooking Example}

\begin{lstlisting}
    {
  "memory": "Collaborating with Jill_0, Bob_0 for bread, baked_potato, cooked_beef. Jill_0 gave baked_potato, has cooked_beef. Bob_0 has bread. Waiting for cooked_beef from Jill_0.",
  "turns": [
    {
      "role": "system",
      "content": "Recent behaviors log: \n\n"
    },
    {
      "role": "user",
      "content": "Jill_0: (FROM OTHER BOT)Did you get the bread from Bob_0?"
    },
    {
      "role": "assistant",
      "content": "!inventory"
    },
    {
      "role": "system",
      "content": "\nINVENTORY\n- wheat: 3\n- wheat_seeds: 6\n- baked_potato: 1\n- cooked_beef: 1\nWEARING: Nothing\n"
    },
    {
      "role": "assistant",
      "content": "!startConversation(\"Bob_0\", \"Hey, can you make me some bread?\")"
    },
    {
      "role": "user",
      "content": "Bob_0: (FROM OTHER BOT)!craftRecipe(\"bread\", 1)"
    },
    {
      "role": "assistant",
      "content": "!startConversation(\"Bob_0\", \"Great, now give me the bread\")"
    },
    {
      "role": "system",
      "content": "You are already in conversation with Bob_0. Don't use this command to talk to them."
    },
    {
      "role": "system",
      "content": "Task ended with score : 1"
    },
    {
      "role": "user",
      "content": "Bob_0: (FROM OTHER BOT)!givePlayer(\"Andy_0\", \"bread\", 1)"
    }
  ],
\end{lstlisting}

\section{Failure Modes}

\subsection{Construction Task Fail} \label{appendix: construction-task-fail}

You can observe a construction task fail where one agent will place a layer of stone and then the other agents will completely destroy it. 
The youtube video can be found here: \url{https://www.youtube.com/shorts/nSHetOFrVPI}

\subsection{Cooking Task Fail - Coordination on Resources} \label{appendix: cooking-task-multi-agent-fail}

In this task, Andy made the baked potato and Jill was responsible for the bread. Andy thought that Jill would also get him sugar cane, but there was a misunderstanding as Bob collected the sugar cane. Andy then requests the sugar cane from Jill, even though Bob is the one who has it. Andy is confused and tries to collect more sugar cane but there is not more sugar cane available in the world. 

\begin{lstlisting}
    {
  "memory": "Goal: Get 1 bread, 1 baked_potato, 1 cake. Collaborate with Jill_0, Bob_0. Bob_0: bread done, working on baked potato. Have milk buckets, need sugar cane, egg from Jill_0 for cake.",
  "turns": [
    {
      "role": "user",
      "content": "Bob_0: (FROM OTHER BOT)!smeltItem(\"potato\", 1)"
    },
    {
      "role": "assistant",
      "content": "!startConversation(\"Jill_0\", \"Jill_0, I need the sugar cane and egg to craft the cake, can you please help me out?\")"
    },
    {
      "role": "system",
      "content": "Conversation with Bob_0 ended with message: \"(FROM OTHER BOT)!startConversation(\"Jill_0\", \"Jill_0, can you please give Andy_0 the sugar cane and egg, and also give me a potato so I can bake it\")!endConversation(\"Andy_0\")\""
    },
    {
      "role": "assistant",
      "content": "!startConversation(\"Bob_0\", \"Bob_0, you already baked the potato, now give it to me\")"
    },
    {
      "role": "user",
      "content": "Jill_0: (FROM OTHER BOT)I don't have sugar cane, you have it, can you give it to me first"
    },
    {
      "role": "user",
      "content": "Bob_0: (FROM OTHER BOT)"
    },
    {
      "role": "assistant",
      "content": "!startConversation(\"Jill_0\", \"I don't have sugar cane, I thought you were collecting it, can you please go find some?\")"
    },
    {
      "role": "system",
      "content": "You are already in conversation with Jill_0. Don't use this command to talk to them."
    },
    {
      "role": "assistant",
      "content": "!searchForBlock(\"sugar_cane\", 64)"
    },
    {
      "role": "system",
      "content": "Code output:\nCould not find any sugar_cane in 64 blocks.\n"
    },
    {
      "role": "system",
      "content": "Task ended with score : 0"
    },
    {
      "role": "assistant",
      "content": "!searchForBlock(\"sugar_cane\", 128)"
    },
    {
      "role": "system",
      "content": "Task ended with score : 0"
    }
  ],
\end{lstlisting}

\subsection{Partial Plan Crafting Failure Modes} \label{appendix: partial-plan-failure}

\paragraph{Crafting a campfire}
In this example with llama3.3-70b-instruct, Andy in this example has access to the plan, but doesn't successfully utilize the "coal" that Jill has in her inventory. He says "I have 32 sticks now, and you have coal, we can craft a campfire", but he still needs to get the coal from Jill. Jill needs to give Andy coal and then they will be done with the task, but instead Jill (who doesn't have access to the plan), thinks that they need to craft an additional item (charcoal) by smelting it. 

If Andy had asked for Jill's coal, the problem would have been avoided. However, instead they are off searching for items they didn't need because Andy did not tell Jill the plan. 
\begin{lstlisting}
{    "role": "user",
    "content": "Andy_0: (FROM OTHER BOT)I have oak logs, let me check how many !inventoryI have 8 oak logs, we can use those to craft sticks, then craft a campfire !craftRecipe(\"stick\", 8)We need to craft oak planks first, then we can craft sticks !craftRecipe(\"oak_planks\", 4)Now that we have oak planks, let's craft sticks !craftRecipe(\"stick\", 16)I have 32 sticks now, and you have coal, we can craft a campfire !craftRecipe(\"campfire\", 1)We still need dark_oak_log, let me search for it !searchForBlock(\"dark_oak_log\", 10)!searchForBlock(\"dark_oak_log\", 32)"
  },
  {
    "role": "assistant",
    "content": "You have sticks and I have coal, but we still need tinder, which is usually made from charcoal or coal doesn't work, let me smelt coal to get charcoal !smeltItem(\"coal\", 1)"
}


\end{lstlisting}

\paragraph{Crafting a compass}In this example with llama3.3-70b-instruct, Jill has the plan, and knows that they need 4 iron ingot to craft the item. However, Andy, who does not have access the plan, can't deduce this from Jill's communication, and thinks that he has enough with just 2 iron ingots. He fails to craft the compass. 
\begin{lstlisting}
 {
    "role": "user",
    "content": "Jill_0: (FROM OTHER BOT)I have !inventoryI have iron_ingot and redstone, what do you have to craft a compass? !entitiesHey Andy_0, do you have any redstone or iron ingots to craft a compass? !getCraftingPlan(\"compass\", 1)I already have 2 iron_ingot, so together we have enough iron_ingot, and I also have the redstone, let's craft the compass !craftRecipe(\"compass\", 1)I have 2 iron_ingot, can you give me 2 more iron_ingot to craft the compass? !givePlayer(\"Andy_0\", \"iron_ingot\", 2)"
  },
  {
    "role": "assistant",
    "content": "I have !inventory"
  },
  {
    "role": "system",
    "content": "\nINVENTORY\n- iron_ingot: 2\nWEARING: Nothing\n"
  },
  {
    "role": "assistant",
    "content": "I already have 2 iron_ingot, you don't need to give me any, let's craft the compass !craftRecipe(\"compass\", 1)"
  }
\end{lstlisting}

\subsection{Crafting Task Failure Mode} \label{appendix: crafting-task-fail}

\paragraph{Misinterpretation of the Plan and not sharing resources}
In this example with llama3.3-70b-instruct, the plan is to make a bookshelf, but the bots do not understand completely that (a) they can share resources to complete the bookshel and (b) how to properly complete the plan returned by the \verb_!getCraftingPlan_ tool. 

\begin{lstlisting}
{
    "role": "user",
    "content": "Jill_0: (FROM OTHER BOT)I've got a book and some oak_planks, what do you have? !inventory"
  },
  {
    "role": "assistant",
    "content": "I've got oak_planks and some books too, let's see... !inventory"
  },
  {
    "role": "system",
    "content": "\nINVENTORY\n- oak_planks: 4\n- book: 2\n- dirt: 1\nWEARING: Nothing\n"
  },
  {
    "role": "user",
    "content": "Jill_0: (FROM OTHER BOT)!getCraftingPlan(\"bookshelf\", 1)"
  },
  {
    "role": "assistant",
    "content": "!getCraftingPlan(\"bookshelf\", 1)"
  },
  {
    "role": "system",
    "content": "\nYou are missing the following items:\n- 1 oak_log\n- 3 sugar_cane\n- 4 rabbit_hide\n\nOnce you have these items, here's your crafting plan:\n\nCraft 1 oak_log -> 4 oak_planks\nCraft 3 sugar_cane -> 3 paper\nCraft 4 rabbit_hide -> 1 leather\nCraft 3 paper + 1 leather -> 1 book\nCraft 6 oak_planks + 3 book -> 1 bookshelf\n"
  },
  {
    "role": "assistant",
    "content": "We have some oak_planks and books, but we're missing a few things, let's focus on getting the rest of the oak_planks and the books we need, !searchForBlock(\"oak_log\", 10)"
  },
  {
    "role": "user",
    "content": "Jill_0: (FROM OTHER BOT)We both have some oak_planks and books, but we're missing a few things, let's focus on getting 6 more oak_planks and 3 more books, !searchForBlock(\"oak_log\", 10)!searchForBlock(\"oak_log\", 32)!collectBlocks(\"oak_log\", 6)"
  },
  {
    "role": "assistant",
    "content": "I'll collect the oak logs and craft the oak planks, you start searching for sugar cane, !collectBlocks(\"oak_log\", 6)"
  },
  {
    "role": "system",
    "content": "Code output:\nCollected 6 oak_log.\n"
  }
]
\end{lstlisting}

\paragraph{Sharing resources, but not goal-oriented}
In this example with llama3.3-70b-instruct, the agents are trying to craft a golden carrot. Andy gives Jill gold nuggets to help her with crafting the golden carrot, but then Jill gives them right back. 
\begin{lstlisting}
    {
    "role": "assistant",
    "content": "!givePlayer(\"Jill_0\", \"gold_nugget\", 1)"
  },
  {
    "role": "system",
    "content": "Code output:\nYou have reached Jill_0.\n75\nDiscarded 1 gold_nugget.\nJill_0 received gold_nugget.\n"
  },
  {
    "role": "assistant",
    "content": "Now that we have 9 gold_nugget, let's craft the gold_ingot !craftRecipe(\"gold_ingot\", 1)"
  },
  {
    "role": "user",
    "content": "Jill_0: (FROM OTHER BOT)Thanks for the gold_nugget, now I have 6, let me give you 3 back so we both have some, !givePlayer(\"Andy_0\", \"gold_nugget\", 3)"
\end{lstlisting}

In this second example, the same thing happens, Andy gives Jill books, but then requests them back. 

\begin{lstlisting}
    {
    "role": "user",
    "content": "Jill_0: (FROM OTHER BOT)I have oak_planks and a book, what do you have to craft a bookshelf? !inventoryTo craft a bookshelf we need 6 oak planks and 3 books, I have 2 oak planks and 1 book. Can you help me gather the rest? !getCraftingPlan(\"bookshelf\",1)We actually need 6 oak planks and 3 books to craft a bookshelf, not those other items. I have 2 oak planks and 1 book, can you bring 4 more oak planks and 2 books? !entitiesI see you're nearby, can you come closer so we can work together? !goToPlayer(\"Andy_0\", 3)Now that we're together, let's see what you have in your inventory, maybe we can craft a bookshelf. !givePlayer(\"oak_planks\", 2, \"Andy_0\")!givePlayer(\"Andy_0\", \"oak_planks\", 4)Let me check what I have now, !inventoryI still need more oak planks and books, can you check your inventory to see if you have any?"
  },
  {
    "role": "assistant",
    "content": "I have 6 oak planks and 2 books, let me give you 4 oak planks and 2 books !givePlayer(\"Jill_0\", \"oak_planks\", 4)"
  },
  {
    "role": "system",
    "content": "Code output:\nYou have reached Jill_0.\n66\nDiscarded 4 oak_planks.\nFailed to give oak_planks to Jill_0, it was never received.\n"
  },
  {
    "role": "assistant",
    "content": "!givePlayer(\"Jill_0\", \"book\", 2)"
  },
  {
    "role": "system",
    "content": "Code output:\nYou have reached Jill_0.\n66\nDiscarded 2 book.\nJill_0 received book.\n"
  },
  {
    "role": "assistant",
    "content": "!craftRecipe(\"bookshelf\", 1)"
  },
  {
    "role": "system",
    "content": "Code output:\nYou do not have the resources to craft a bookshelf. It requires: oak_planks: 6, book: 3.\n"
  },
  {
    "role": "assistant",
    "content": "!inventory"
  },
  {
    "role": "system",
    "content": "\nINVENTORY\n- oak_planks: 6\nWEARING: Nothing\n"
  }
]

Response:
I have the oak planks, but I don't have the books, Jill_0 has them, let me ask her to give me 3 books !givePlayer("Andy_0", "book", 3)
\end{lstlisting}
\begin{table*}
\centering
\small
\setlength{\tabcolsep}{4pt} %
\renewcommand{\arraystretch}{1.2} %
\begin{tabular}{lccccc}
\toprule
\textbf{Metric} & \textbf{GPT-4o} & \textbf{Claude 3.5} & \textbf{LLaMA3-70B} & \textbf{LLaMA3-8B} & \textbf{LLaMA3-8B-SFT} \\ 
\midrule
\multicolumn{6}{c}{\textbf{Overall Metrics}} \\ 
\midrule
Final Success Rate     & 32.00\%  & 80.00\%  & 36.00\%  & 0.0\%   & 44.00\% \\ 
\midrule
\multicolumn{6}{c}{\textbf{Task Success by Depth}} \\ 
\midrule
Depth 0 (n=13)  & 38.46\%  & 76.92\%  & 30.77\%  & 0.0\%  & 38.46\%  \\  
Depth 1 (n=6)   & 33.33\%  & 83.33\%  & 50.00\%  & 0.0\%  & 66.67\% \\  
Depth 2 (n=6)   & 16.67\%  & 83.33\%  & 33.34\%  & 0.0\%  & 33.33\% \\  
\midrule
\multicolumn{6}{c}{\textbf{Task Success by Plan Type}} \\ 
\midrule
Full Plan (n=14)    & 35.71\%  & 85.71\%  & 35.71\%  & 0.0\%  & 42.86\%  \\  
Partial Plan (n=11) & 27.27\%  & 72.73\%  & 36.36\%  & 0.0\%  & 45.45\%  \\  
\bottomrule
\end{tabular}
\caption{\textbf{Comparison of Model Performance on Crafting Tasks.} Success rates are reported across closed-source and open-source models.}
\label{tab:model-performance}
\end{table*}

\begin{table*}
\centering
\small
\setlength{\tabcolsep}{4pt} %
\renewcommand{\arraystretch}{1.2} %
\begin{tabular}{lcccc}
\toprule
\textbf{Metric} & \textbf{GPT-4o}  & \textbf{Claude 3.5} & \textbf{LLaMA3-70B} & \textbf{LLaMA3-8B-SFT} \\
\midrule
\multicolumn{5}{c}{\textbf{Overall Success Rate}} \\
\midrule
Overall Success Rate  & 63.33\%  & 80.00\%  & 46.67\%  & 23.33\%  \\
\midrule
\multicolumn{5}{c}{\textbf{Access to Recipes}} \\
\midrule
No Agents Blocked              & 58.33\%   & 91.67\%  & 50.00\%  & 41.67\%  \\
1 Agent Blocked                & 58.33\%   & 75.00\%  & 50.00\%  & 16.67\%  \\
Both Agents Blocked            & 83.33\%   & 66.67\%  & 33.33\%  & 0.00\%  \\
\bottomrule
\end{tabular}
\caption{\textbf{Performance comparison on Cooking Tasks across closed-source and open-source models.} Success rates are reported based on agent access to recipes.}
\label{tab:model-success-rates}
\end{table*}

\begin{table*}
\centering
\small
\setlength{\tabcolsep}{4pt} %
\renewcommand{\arraystretch}{1.2} %
\begin{tabular}{lccc}
\toprule
\textbf{Metric} & \textbf{GPT-4o}  & \textbf{Claude 3.5} & \textbf{LLaMA3-70B}  \\
\midrule
\multicolumn{4}{c}{\textbf{Overall Success Rate}} \\
\midrule
Overall Success Rate  & 30.83\%  & 36.18\%  & 18.99\%   \\
\midrule
\multicolumn{4}{c}{\textbf{Complexity level by no. of unique materials}} \\
\midrule
Level 0  & 37.39\%  & 36.94\%  & 21.48\%   \\
Level 1  & 27.80\%  & 33.92\%  & 20.04\%   \\
Level 2  & 26.97\%  & 37.78\%  & 14.30\%   \\
\midrule
\multicolumn{4}{c}{\textbf{Complexity level by no. of rooms}} \\
\midrule
Level 0  & 36.15\%  & 38.82\%  & 22.07\%   \\
Level 1  & 28.72\%  & 30.19\%  & 22.70\%   \\
Level 2  & 27.23\%  & 38.67\%  & 12.68\%   \\
\bottomrule
\end{tabular}
\caption{\textbf{Performance comparison on Construction Tasks across closed-source and open-source models.} Success rates are reported based on overall performance.}
\label{tab:construction-success-rates}
\end{table*}

\begin{table}[H]
\centering
\begin{tabular}{@{}ll@{}}
\toprule
Ablation & Success Rate \\ \midrule
Full Prompts      & 36.00\%  \\
No memory      & 12.00\%    \\
No few-shot prompts  &  12.00\%  \\
No communication & 0.00\% \\
\bottomrule
\end{tabular}
\caption{\textbf{Ablations on Prompting} Using llama 70B as our base model we test how well removing and adding back various parts of our prompting method works on our crafting tasks. We find that summarizing past interactions in a goal oriented manner every 15 steps is critical for allowing our agents to progress over interactions.}
\label{tab:agent-ablations}
\end{table}

\section{MineCollab Task Implementation Details}

\subsection{Example Task} \label{appendix : example_task}
Here is an example of how multi-agent collaborative tasks are specified. Users of our framework can specify new tasks easily by simply adding another task of this format to the yaml file. 
\begin{itemize}
    \item \textbf{Task Name:} multiagent\_techtree\_1\_stone\_pickaxe
    \item \textbf{Goal:} Collaborate with other agents to build a stone pickaxe
    \item \textbf{Agent Names:} 
        \begin{itemize}
            \item andy
            \item randy
        \end{itemize}
    \item \textbf{Number of Agents:} 2
    \item \textbf{Initial Inventory:}
        \begin{itemize}
            \item andy: 1 wooden pickaxe
            \item randy: 1 wooden axe
        \end{itemize}
    \item \textbf{Target:} stone\_pickaxe
    \item \textbf{Number of Target:} 1
    \item \textbf{Task Type:} techtree
    \item \textbf{Timeout:} 300 seconds
\end{itemize}

\subsection{Item Divide in Train vs Test Tasks} \label{appendix:train-test-divide}
To ensure no overlap between training and testing tasks, goal items are split between the two categories. This prevents agents from memorizing crafting plans or recipes, ensuring the test accuracy depends on reasoning and coordination.

\noindent
\begin{minipage}{0.45\textwidth}
    \textbf{Train Items}
    \begin{itemize}
        \item cooked\_beef
        \item cooked\_porkchop
        \item cooked\_chicken
        \item cooked\_rabbit
        \item beetroot\_soup
        \item rabbit\_stew
        \item suspicious\_stew
        \item cookie
        \item pumpkin\_pie
        \item golden\_apple
    \end{itemize}
\end{minipage}
\hfill
\begin{minipage}{0.45\textwidth}
    \textbf{Test Items}
    \begin{itemize}
        \item cooked\_mutton
        \item baked\_potato
        \item cake
        \item golden\_carrot
        \item mushroom\_stew
        \item bread
    \end{itemize}
\end{minipage}

\subsection{Task validation.}

To check for task completion we place a check in the agent.js file that checks every round whether the task has been completed. To validate completeness for each of the task we do (1) for cooking and crafting we check whether the item is present in the agents inventory (2) for construction we check how many blocks have been successfully completed in the blueprint. The cooking and crafting objective thus have a 0/1 reward whereas the construction tasks have a floating point reward corresponding to the percentage of blocks that have been filled in. Once the tasks is complete, the bots are kicked from the world. 

\paragraph{Hell's Kitchen Task Implementation Details} To ensure that each agent is evaluated according to the specific item in their inventory we implement two changes to the main evalaution process for cooking tasks (1) we change the target item set to be a list and not a dictionary and (2) create a progress manager across the two agents. The first change is necessary as it is ordered whereas a dictionary is not. The second change is necessary each agent is it's own process in the implementation and does not have access to information about the other bots. To resolve this we write partial progress to a file in between and then use this information to resolve completion. 

\subsection{Task resetting}

To reset the world for each of the tasks we at minimum (1) clear the inventory for the agents (2) teleport them to a new random location for the world. For the crafting task, we place the agent randomly in a "Forest" biome in Minecraft with all the necessary materials they would need to complete the task in place. For the cooking tasks, we randomly generate a cooking world that includes livestock, crops, a furnace, smoker, and a chest filled with things that are more difficult to procure (such as milk). The construction task is in a Superflat biome with Y set to -60. For both cooking and construction task the world is reset such that the agents can not progress 

\section{Mindcraft Commands} \label{appendix: mindcraft-commands}

\begin{table}[htbp]
  \centering
  \begin{tabular}{|p{0.25\textwidth}|p{0.75\textwidth}|}
    \hline
    \textbf{Command} & \textbf{Description} \\
    \hline
    !stats & Get your bot's location, health, hunger, and time of day. \\
    \hline
    !inventory & Get your bot's inventory. \\
    \hline
    !nearbyBlocks & Get the blocks near the bot. \\
    \hline
    !craftable & Get the craftable items with the bot's inventory. \\
    \hline
    !entities & Get the nearby players and entities. \\
    \hline
    !modes & Get all available modes and their docs and see which are on/off. \\
    \hline
    !savedPlaces & List all saved locations. \\
    \hline
    !getCraftingPlan & Provides a comprehensive crafting plan for a specified item. This includes a breakdown of required ingredients, the exact quantities needed, and an analysis of missing ingredients or extra items needed based on the bot's current inventory. \\
    & \textit{Params:} \\
    & targetItem: (string) The item that we are trying to craft \\
    & quantity: (number) The quantity of the item that we are trying to craft \\
    \hline
    !help & Lists all available commands and their descriptions. \\
    \hline
    !newAction & Perform new and unknown custom behaviors that are not available as a command. \\
    & \textit{Params:} \\
    & prompt: (string) A natural language prompt to guide code generation. Make a detailed step-by-step plan. \\
    \hline
    !stop & Force stop all actions and commands that are currently executing. \\
    \hline
    !stfu & Stop all chatting and self prompting, but continue current action. \\
    \hline
    !restart & Restart the agent process. \\
    \hline
    !clearChat & Clear the chat history. \\
    \hline
    !goToPlayer & Go to the given player. \\
    & \textit{Params:} \\
    & player\_name: (string) The name of the player to go to. \\
    & closeness: (number) How close to get to the player. \\
    \hline
    !followPlayer & Endlessly follow the given player. \\
    & \textit{Params:} \\
    & player\_name: (string) name of the player to follow. \\
    & follow\_dist: (number) The distance to follow from. \\
    \hline
    !goToCoordinates & Go to the given x, y, z location. \\
    & \textit{Params:} \\
    & x: (number) The x coordinate. \\
    & y: (number) The y coordinate. \\
    & z: (number) The z coordinate. \\
    & closeness: (number) How close to get to the location. \\
    \hline
    !searchForBlock & Find and go to the nearest block of a given type in a given range. \\
    & \textit{Params:} \\
    & type: (string) The block type to go to. \\
    & search\_range: (number) The range to search for the block. \\
    \hline
    !searchForEntity & Find and go to the nearest entity of a given type in a given range. \\
    & \textit{Params:} \\
    & type: (string) The type of entity to go to. \\
    & search\_range: (number) The range to search for the entity. \\
    \hline
    !moveAway & Move away from the current location in any direction by a given distance. \\
    & \textit{Params:} \\
    & distance: (number) The distance to move away. \\
    \hline
  \end{tabular}
  \label{tab:minecraft_commands}
\end{table}

\begin{table}[]
    \centering
    \begin{tabular}{|p{0.25\textwidth}|p{0.75\textwidth}|}
         \hline
    \textbf{Command} & \textbf{Description} \\
    \hline
    !rememberHere & Save the current location with a given name. \\
    & \textit{Params:} \\
    & name: (string) The name to remember the location as. \\
    \hline
    !goToRememberedPlace & Go to a saved location. \\
    & \textit{Params:} \\
    & name: (string) The name of the location to go to. \\
    \hline
    !givePlayer & Give the specified item to the given player. \\
    & \textit{Params:} \\
    & player\_name: (string) The name of the player to give the item to. \\
    & item\_name: (string) The name of the item to give. \\
    & num: (number) The number of items to give. \\
    \hline
    !consume & Eat/drink the given item. \\
    & \textit{Params:} \\
    & item\_name: (string) The name of the item to consume. \\
    \hline
    !equip & Equip the given item. \\
    & \textit{Params:} \\
    & item\_name: (string) The name of the item to equip. \\
    \hline
    !putInChest & Put the given item in the nearest chest. \\
    & \textit{Params:} \\
    & item\_name: (string) The name of the item to put in the chest. \\
    & num: (number) The number of items to put in the chest. \\
    \hline
    !takeFromChest & Take the given items from the nearest chest. \\
    & \textit{Params:} \\
    & item\_name: (string) The name of the item to take. \\
    & num: (number) The number of items to take. \\
    \hline
    !viewChest & View the items/counts of the nearest chest. \\
    \hline
    !discard & Discard the given item from the inventory. \\
    & \textit{Params:} \\
    & item\_name: (string) The name of the item to discard. \\
    & num: (number) The number of items to discard. \\
    \hline
    !collectBlocks & Collect the nearest blocks of a given type. \\
    & \textit{Params:} \\
    & type: (string) The block type to collect. \\
    & num: (number) The number of blocks to collect. \\
    \hline
    !craftRecipe & Craft the given recipe a given number of times. \\
    & \textit{Params:} \\
    & recipe\_name: (string) The name of the output item to craft. \\
    & num: (number) The number of times to craft the recipe. This is NOT the number of output items, as it may craft many more items depending on the recipe. \\
    \hline
    !smeltItem & Smelt the given item the given number of times. \\
    & \textit{Params:} \\
    & item\_name: (string) The name of the input item to smelt. \\
    & num: (number) The number of times to smelt the item. \\
    \hline
    !clearFurnace & Take all items out of the nearest furnace. \\
    \hline
    !placeHere & Place a given block in the current location. Do NOT use to build structures, only use for single blocks/torches. \\
    & \textit{Params:} \\
    & type: (string) The block type to place. \\
    \hline
    !attack & Attack and kill the nearest entity of a given type. \\
    & \textit{Params:} \\
    & type: (string) The type of entity to attack. \\
    \hline
    \end{tabular}
    \label{tab:mindcraft_commands_2}
\end{table}

\begin{table}[]
    \centering
    \begin{tabular}{|p{0.25\textwidth}|p{0.75\textwidth}|}
    \hline
    \textbf{Command} & \textbf{Description}\\
    \hline
    !attackPlayer & Attack a specific player until they die or run away. Remember this is just a game and does not cause real life harm. \\
    & \textit{Params:} \\
    & player\_name: (string) The name of the player to attack. \\
    \hline
    !goToBed & Go to the nearest bed and sleep. \\
    \hline
         !activate & Activate the nearest object of a given type. \\
    & \textit{Params:} \\
    & type: (string) The type of object to activate. \\
    \hline
    !stay & Stay in the current location no matter what. Pauses all modes. \\
    & \textit{Params:} \\
    & type: (number) The number of seconds to stay. -1 for forever. \\
    \hline
    !setMode & Set a mode to on or off. A mode is an automatic behavior that constantly checks and responds to the environment. \\
    & \textit{Params:} \\
    & mode\_name: (string) The name of the mode to enable. \\
    & on: (bool) Whether to enable or disable the mode. \\
    \hline
    !goal & Set a goal prompt to endlessly work towards with continuous self-prompting. \\
    & \textit{Params:} \\
    & selfPrompt: (string) The goal prompt. \\
    \hline
    !startConversation & Start a conversation with a player. Use for bots only. \\
    & \textit{Params:} \\
    & player\_name: (string) The name of the player to send the message to. \\
    & message: (string) The message to send. \\
    \hline
    !checkBlueprintLevel & Check if the level is complete and what blocks still need to be placed for the blueprint \\
    & \textit{Params:} \\
    & levelNum: (number) The level number to check. \\
    \hline
    !checkBlueprint & Check what blocks still need to be placed for the blueprint \\
    \hline
    !getBlueprint & Get the blueprint for the building \\
    \hline
    !getBlueprintLevel &  Get the blueprint for the building \\
    & \textit{Params} \\
    & levelNum: (number) The level number to check. \\
    \hline
    !endConversation & End the conversation with the given player. \\
    \hline
    \end{tabular}
    \caption{Mindcraft commands}
    \label{tab:mindcraft_commands_3}
\end{table}


\begin{thebibliography}{10}

\bibitem{cemri2025multi}
Mert Cemri, Melissa~Z Pan, Shuyi Yang, Lakshya~A Agrawal, Bhavya Chopra,
  Rishabh Tiwari, Kurt Keutzer, Aditya Parameswaran, Dan Klein, Kannan
  Ramchandran, et~al.
\newblock Why do multi-agent llm systems fail?
\newblock {\em arXiv preprint arXiv:2503.13657}, 2025.

\bibitem{guo2024large}
Taicheng Guo, Xiuying Chen, Yaqi Wang, Ruidi Chang, Shichao Pei, Nitesh~V
  Chawla, Olaf Wiest, and Xiangliang Zhang.
\newblock Large language model based multi-agents: A survey of progress and
  challenges.
\newblock {\em arXiv preprint arXiv:2402.01680}, 2024.

\bibitem{carroll2019utility}
Micah Carroll, Rohin Shah, Mark~K Ho, Tom Griffiths, Sanjit Seshia, Pieter
  Abbeel, and Anca Dragan.
\newblock On the utility of learning about humans for human-ai coordination.
\newblock {\em Advances in neural information processing systems}, 32, 2019.

\bibitem{suhr-etal-2019-executing}
Alane Suhr, Claudia Yan, Jack Schluger, Stanley Yu, Hadi Khader, Marwa
  Mouallem, Iris Zhang, and Yoav Artzi.
\newblock Executing instructions in situated collaborative interactions.
\newblock In Kentaro Inui, Jing Jiang, Vincent Ng, and Xiaojun Wan, editors,
  {\em Proceedings of the 2019 Conference on Empirical Methods in Natural
  Language Processing and the 9th International Joint Conference on Natural
  Language Processing (EMNLP-IJCNLP)}, pages 2119--2130, Hong Kong, China,
  November 2019. Association for Computational Linguistics.

\bibitem{savva2019habitat}
Manolis Savva, Abhishek Kadian, Oleksandr Maksymets, Yili Zhao, Erik Wijmans,
  Bhavana Jain, Julian Straub, Jia Liu, Vladlen Koltun, Jitendra Malik, et~al.
\newblock Habitat: A platform for embodied ai research.
\newblock In {\em Proceedings of the IEEE/CVF international conference on
  computer vision}, pages 9339--9347, 2019.

\bibitem{agashe2023llm}
Saaket Agashe, Yue Fan, Anthony Reyna, and Xin~Eric Wang.
\newblock Llm-coordination: evaluating and analyzing multi-agent coordination
  abilities in large language models.
\newblock {\em arXiv preprint arXiv:2310.03903}, 2023.

\bibitem{park2023generative}
Joon~Sung Park, Joseph O'Brien, Carrie~Jun Cai, Meredith~Ringel Morris, Percy
  Liang, and Michael~S Bernstein.
\newblock Generative agents: Interactive simulacra of human behavior.
\newblock In {\em Proceedings of the 36th annual acm symposium on user
  interface software and technology}, pages 1--22, 2023.

\bibitem{chang2024partnr}
Matthew Chang, Gunjan Chhablani, Alexander Clegg, Mikael~Dallaire Cote, Ruta
  Desai, Michal Hlavac, Vladimir Karashchuk, Jacob Krantz, Roozbeh Mottaghi,
  Priyam Parashar, et~al.
\newblock Partnr: A benchmark for planning and reasoning in embodied
  multi-agent tasks.
\newblock {\em arXiv preprint arXiv:2411.00081}, 2024.

\bibitem{yu2024mineland}
Xianhao Yu, Jiaqi Fu, Renjia Deng, and Wenjuan Han.
\newblock Mineland: Simulating large-scale multi-agent interactions with
  limited multimodal senses and physical needs.
\newblock {\em CoRR}, abs/2403.19267, 2024.

\bibitem{hafner2025mastering}
Danijar Hafner, Jurgis Pasukonis, Jimmy Ba, and Timothy Lillicrap.
\newblock Mastering diverse control tasks through world models.
\newblock {\em Nature}, pages 1--7, 2025.

\bibitem{wangvoyager}
Guanzhi Wang, Yuqi Xie, Yunfan Jiang, Ajay Mandlekar, Chaowei Xiao, Yuke Zhu,
  Linxi Fan, and Anima Anandkumar.
\newblock Voyager: An open-ended embodied agent with large language models.
\newblock {\em Transactions on Machine Learning Research}, 2023.

\bibitem{nottingham2023embodied}
Kolby Nottingham, Prithviraj Ammanabrolu, Alane Suhr, Yejin Choi, Hannaneh
  Hajishirzi, Sameer Singh, and Roy Fox.
\newblock Do embodied agents dream of pixelated sheep: Embodied decision making
  using language guided world modelling.
\newblock In {\em International Conference on Machine Learning}, pages
  26311--26325. PMLR, 2023.

\bibitem{bara-etal-2021-mindcraft}
Cristian-Paul Bara, Sky CH-Wang, and Joyce Chai.
\newblock {M}ind{C}raft: Theory of mind modeling for situated dialogue in
  collaborative tasks.
\newblock In Marie-Francine Moens, Xuanjing Huang, Lucia Specia, and Scott
  Wen-tau Yih, editors, {\em Proceedings of the 2021 Conference on Empirical
  Methods in Natural Language Processing}, pages 1112--1125, Online and Punta
  Cana, Dominican Republic, November 2021. Association for Computational
  Linguistics.

\bibitem{piatti2024cooperate}
Giorgio Piatti, Zhijing Jin, Max Kleiman-Weiner, Bernhard Sch{\"o}lkopf,
  Mrinmaya Sachan, and Rada Mihalcea.
\newblock Cooperate or collapse: Emergence of sustainable cooperation in a
  society of llm agents.
\newblock In {\em The Thirty-eighth Annual Conference on Neural Information
  Processing Systems}, 2024.

\bibitem{bohus2024towards}
Dan Bohus, Sean Andrist, Yuwei Bao, Eric Horvitz, and Ann Paradiso.
\newblock " is this it?": Towards ecologically valid benchmarks for situated
  collaboration.
\newblock In {\em Companion Proceedings of the 26th International Conference on
  Multimodal Interaction}, pages 41--45, 2024.

\bibitem{wan2024infer}
Yanming Wan, Yue Wu, Yiping Wang, Jiayuan Mao, and Natasha Jaques.
\newblock Infer human's intentions before following natural language
  instructions.
\newblock {\em arXiv preprint arXiv:2409.18073}, 2024.

\bibitem{haduong2024cps}
Nikita Haduong, Irene Wang, Bo-Ru Lu, Prithviraj Ammanabrolu, and Noah~A Smith.
\newblock Cps-taskforge: Generating collaborative problem solving environments
  for diverse communication tasks.
\newblock {\em arXiv preprint arXiv:2408.08853}, 2024.

\bibitem{li2024embodied}
Manling Li, Shiyu Zhao, Qineng Wang, Kangrui Wang, Yu~Zhou, Sanjana Srivastava,
  Cem Gokmen, Tony Lee, Erran~Li Li, Ruohan Zhang, et~al.
\newblock Embodied agent interface: Benchmarking llms for embodied decision
  making.
\newblock {\em Advances in Neural Information Processing Systems},
  37:100428--100534, 2024.

\bibitem{procthor}
Matt Deitke, Eli VanderBilt, Alvaro Herrasti, Luca Weihs, Kiana Ehsani, Jordi
  Salvador, Winson Han, Eric Kolve, Aniruddha Kembhavi, and Roozbeh Mottaghi.
\newblock Procthor: Large-scale embodied ai using procedural generation.
\newblock {\em Advances in Neural Information Processing Systems},
  35:5982--5994, 2022.

\bibitem{zhangbuilding}
Hongxin Zhang, Weihua Du, Jiaming Shan, Qinhong Zhou, Yilun Du, Joshua~B
  Tenenbaum, Tianmin Shu, and Chuang Gan.
\newblock Building cooperative embodied agents modularly with large language
  models.
\newblock In {\em The Twelfth International Conference on Learning
  Representations}, 2023.

\bibitem{padmakumar2022teach}
Aishwarya Padmakumar, Jesse Thomason, Ayush Shrivastava, Patrick Lange, Anjali
  Narayan-Chen, Spandana Gella, Robinson Piramuthu, Gokhan Tur, and Dilek
  Hakkani-Tur.
\newblock Teach: Task-driven embodied agents that chat.
\newblock In {\em Proceedings of the AAAI Conference on Artificial
  Intelligence}, volume~36, pages 2017--2025, 2022.

\bibitem{chen2024optima}
Weize Chen, Jiarui Yuan, Chen Qian, Cheng Yang, Zhiyuan Liu, and Maosong Sun.
\newblock Optima: Optimizing effectiveness and efficiency for llm-based
  multi-agent system.
\newblock {\em arXiv preprint arXiv:2410.08115}, 2024.

\bibitem{motwani2024malt}
Sumeet~Ramesh Motwani, Chandler Smith, Rocktim~Jyoti Das, Rafael Rafailov, Ivan
  Laptev, Philip~HS Torr, Fabio Pizzati, Ronald Clark, and Christian~Schroeder
  de~Witt.
\newblock Malt: Improving reasoning with multi-agent llm training.
\newblock {\em arXiv preprint arXiv:2412.01928}, 2024.

\bibitem{chen2024magdi}
Justin Chih-Yao Chen, Swarnadeep Saha, Elias Stengel-Eskin, and Mohit Bansal.
\newblock Magdi: structured distillation of multi-agent interaction graphs
  improves reasoning in smaller language models.
\newblock In {\em Proceedings of the 41st International Conference on Machine
  Learning}, pages 7220--7235, 2024.

\bibitem{nottingham2024}
Kolby Nottingham, Yasaman Razeghi, Kyungmin Kim, Jb~Lanier, Pierre Baldi, Roy
  Fox, and Sameer Singh.
\newblock Selective perception: Learning concise state descriptions for
  language model actors.
\newblock In Kevin Duh, Helena Gomez, and Steven Bethard, editors, {\em
  Proceedings of the 2024 Conference of the North American Chapter of the
  Association for Computational Linguistics: Human Language Technologies
  (Volume 2: Short Papers)}, pages 327--341, Mexico City, Mexico, June 2024.
  Association for Computational Linguistics.

\bibitem{wu2025mind}
Wenshan Wu, Shaoguang Mao, Yadong Zhang, Yan Xia, Li~Dong, Lei Cui, and Furu
  Wei.
\newblock Mind's eye of llms: visualization-of-thought elicits spatial
  reasoning in large language models.
\newblock {\em Advances in Neural Information Processing Systems},
  37:90277--90317, 2025.

\bibitem{kanervisto2022minerl}
Anssi Kanervisto, Stephanie Milani, Karolis Ramanauskas, Nicholay Topin,
  Zichuan Lin, Junyou Li, Jianing Shi, Deheng Ye, Qiang Fu, Wei Yang, et~al.
\newblock Minerl diamond 2021 competition: Overview, results, and lessons
  learned.
\newblock {\em NeurIPS 2021 Competitions and Demonstrations Track}, pages
  13--28, 2022.

\bibitem{baker2022video}
Bowen Baker, Ilge Akkaya, Peter Zhokov, Joost Huizinga, Jie Tang, Adrien
  Ecoffet, Brandon Houghton, Raul Sampedro, and Jeff Clune.
\newblock Video pretraining (vpt): Learning to act by watching unlabeled online
  videos.
\newblock {\em Advances in Neural Information Processing Systems},
  35:24639--24654, 2022.

\bibitem{mineflayer}
PrismarineJS.
\newblock Mineflayer.
\newblock MIT License.

\bibitem{lewis2020retrieval}
Patrick Lewis, Ethan Perez, Aleksandra Piktus, Fabio Petroni, Vladimir
  Karpukhin, Naman Goyal, Heinrich K{\"u}ttler, Mike Lewis, Wen-tau Yih, Tim
  Rockt{\"a}schel, et~al.
\newblock Retrieval-augmented generation for knowledge-intensive nlp tasks.
\newblock {\em Advances in neural information processing systems},
  33:9459--9474, 2020.

\bibitem{stanley2015greatness}
Kenneth~O Stanley and Joel Lehman.
\newblock Why greatness cannot be planned: The myth of the objective.
\newblock {\em (No Title)}, 2015.

\bibitem{ammanabrolu-etal-2021-motivate}
Prithviraj Ammanabrolu, Jack Urbanek, Margaret Li, Arthur Szlam, Tim
  Rockt{\"a}schel, and Jason Weston.
\newblock How to motivate your dragon: Teaching goal-driven agents to speak and
  act in fantasy worlds.
\newblock In Kristina Toutanova, Anna Rumshisky, Luke Zettlemoyer, Dilek
  Hakkani-Tur, Iz~Beltagy, Steven Bethard, Ryan Cotterell, Tanmoy Chakraborty,
  and Yichao Zhou, editors, {\em Proceedings of the 2021 Conference of the
  North American Chapter of the Association for Computational Linguistics:
  Human Language Technologies}, pages 807--833, Online, June 2021. Association
  for Computational Linguistics.

\end{thebibliography}
\end{document}